\documentclass[aps,prl,superscriptaddress,twocolumn]{revtex4}

\usepackage{amssymb}
\usepackage{graphicx}
\usepackage{amsmath}
\usepackage{amsthm}
\usepackage{amsfonts}
\usepackage{txfontsb}
\usepackage{color}
\usepackage{bm}
\usepackage{bbm}
\usepackage{url}
\usepackage{xcolor}
\usepackage[driverfallback=dvipdfm]{hyperref}
\hypersetup{pdfpagemode=FullScreen,colorlinks=true,breaklinks,urlcolor=blue,linkcolor=blue,citecolor=blue}

\usepackage{multirow}
\usepackage{physics}

\begin{document}

\global\long\def\id{\mathbbm{1}}
\global\long\def\ui{\mathbbm{i}}
\global\long\def\ud{\mathrm{d}}

\title{Measuring Non-Hermitian Topological Invariants Directly from Quench Dynamics}

\author{Xiao-Dong Lin}
\affiliation{School of Physics and Institute for Quantum Science and Engineering, Huazhong University of Science and Technology, Wuhan 430074, China}
\affiliation{Hefei National Research Center for Physical Sciences at the Microscale and Department of Modern Physics, University of Science and Technology of China, Hefei 230026, China}
\author{Long Zhang}
\email{lzhangphys@hust.edu.cn}
\affiliation{School of Physics and Institute for Quantum Science and Engineering, Huazhong University of Science and Technology, Wuhan 430074, China}
\affiliation{Hefei National Laboratory, Hefei 230088, China}

\begin{abstract} 
While non-Hermitian (NH) topological phases and phenomena have been observed across various quantum systems, directly measuring NH topological invariants remains a significant challenge. In this study, we present a generic and unified framework for the direct measurement of various NH topological invariants in odd-dimensional systems through quench dynamics. We demonstrate that in one-dimensional (1D) NH systems with sublattice symmetry, the line-gap winding number and point-gap braiding degree can be extracted from the winding patterns of a dynamically constructed field based on post-quench spin textures. Specifically, line-gap topology is characterized by integer-valued winding, whereas point-gap complex-band braiding is revealed by half-integer or integer winding with abrupt jumps. We also extend our approach to higher-dimensional winding numbers and non-Bloch topological invariants under open-boundary conditions. Additionally, we propose a practical cold-atom setup to realize and detect 1D NH topological phases, showing that our dynamical measurement scheme is feasible in current experimental settings. This work paves the way for the direct measurement of NH topological invariants in quantum systems.
\end{abstract}

\maketitle

{\em Introduction.---} 
The interplay between non-Hermiticity and topology gives rise to a range of fascinating phenomena that have no Hermitian counterparts~\cite{Bergholtz2021_review,Ding2022_review,Zhang2022_review,Lin2023_review,Okuma2023_review,Banerjee2023_review}, 
such as exceptional points (EPs)~\cite{Ding2022_review,Berry2004,Heiss2012} and non-Hermitian (NH) skin effects~\cite{Zhang2022_review,Lin2023_review,Okuma2023_review,Yao2018a}. 
These phenomena have garnered significant attention and have been extensively explored both theoretically~\cite{Yao2018a,Lee2016,Xu2017,Leykam2017,Yao2018b,Yokomizo2019,Yang2020,Shen2018,Kunst2018,Lieu2018,Yin2018,Gong2018,Kawabata2019a,ZhouH2019,LiuCH2019,Kawabata2019b,Liu2019,LeeCH2019,Luo2019,LeeJY2019,Borgnia2020,Okuma2020,Zhang2020,Sun2021,Denner2021,Nakamura2024,Li2020a,Li2020b,Wojcik2020,LiZ2021,Yang2021,Hu2021,Kawabata2021,Zirnstein2021,Guo2021,Delplace2021,Lv2022,Zhang2022,Zhou2022,Ghosh2022,Guo2023,WangZY2023,Li2023,Hu2024,Wang2024,Xiong2024,Yoshida2024} and experimentally~\cite{Ding2016,Doppler2016,Zhou2018,Tang2020,Ghatak2020,Chen2020,Helbig2020,WangK2021,ZhangX2021,ZhangL2021,Zou2021,Nasari2022,ZhangQ2023,Liang2023,Li2019,Wu2019,Weidemann2020,Xiao2020,Gou2020,Liu2021,Wang2021,ZhangW2021,Su2021,Ren2022,Yu2022,Liang2022,Zhao2023,WuY2023,Cao2023,WangC2024,Xiao2024}. In classifying NH topological bands, two types of complex-energy gaps—line gap and point gap—are typically employed~\cite{Gong2018,Kawabata2019a,ZhouH2019,LiuCH2019,Kawabata2019b}. NH Hamiltonians exhibiting line gaps can be continuously deformed into a Hermitian limit, allowing the topological invariants from Hermitian systems to be naturally extended to characterize line-gap topology~\cite{Shen2018,Kawabata2019a}. Under open boundary conditions, non-Bloch topological invariants are defined within the framework of the generalized Brillouin zone (GBZ) to restore the bulk-boundary correspondence~\cite{Yao2018a,Yao2018b,Yokomizo2019,Yang2020}. In contrast, point-gap topology is intrinsic to NH systems and leads to unconventional boundary behaviors~\cite{Gong2018,LeeJY2019,Borgnia2020,Okuma2020,Zhang2020,Sun2021,Denner2021,Nakamura2024}; notably, the NH skin effects in one dimension arise from the spectral winding associated with point gaps~\cite{Okuma2020,Zhang2020}. Beyond these conventional classifications, braid and knot theories have been developed for NH Hamiltonians with separable bands~\cite{Wojcik2020,LiZ2021,Hu2021}.

Despite the observation of NH topological phenomena across various quantum platforms~\cite{Li2019,Wu2019,Weidemann2020,Xiao2020,Gou2020,Liu2021,Wang2021,ZhangW2021,Su2021,Ren2022,Yu2022,Liang2022,Zhao2023,WuY2023,Cao2023,WangC2024,Xiao2024}, direct measurement of NH topological invariants remains a challenge. 
Here, we propose a generic scheme to measure NH topological invariants through quench-induced nonunitary dynamics. 
Unlike previous works~\cite{Wang2021,ZhangW2021,Su2021,ZhouL2018,Zhu2020,LiT2021,He2023,Nehra2024}, our scheme is capable of characterizing both line-gap and point-gap topology, applicable to one-dimensional (1D) and higher-dimensional systems with sublattice or chiral symmetry, and it accommodates both periodic and open boundary conditions. 
Importantly, all proposed measurements are defined solely within the basis of right eigenvectors, making them straightforward to implement experimentally.

In this Letter, we introduce the winding number $W_d$ of a dynamical field ${\bm g}({\bm k})$, derived from quench measurements of spin textures, to characterize the NH topology in odd dimensions.
Our key findings are as follows: (i)  In 1D systems exhibiting sublattice symmetry (SLS), i.e., $SH({k})S^{-1}= -H({k})$~\cite{Kawabata2019a}, 
the topology of real or imaginary line gaps is characterized by an integer-valued $W_1$ [see Eq.~\eqref{W1}]. 
This result extends seamlessly to higher-dimensional line-gapped systems with SLS, where the topology is defined by $W_d$ [see Eq.~\eqref{Wd}].
(ii) The point-gap topology in 1D systems is identified by a half-integer or integer-valued $W_1$, with abrupt $\pi$ jumps in the winding of ${\bm g}(k)$, 
highlighting the braiding structure of complex-energy bands [see Eq.~\eqref{braid_nu}].
This characterization also applies to higher-dimensional systems, such as NH topological semimetals.
(iii) Open-boundary systems with complex-energy bands can be similarly characterized, 
with all measurements projected onto the GBZ. For systems featuring a real open-boundary spectrum, some modifications are required. 
(iv) We propose a practical cold-atom setup to realize a 1D NH topological system, with detailed studies of its realization and detection using 
$^{40}$K atoms as a specific example.
Additionally, a comprehensive discussion on NH systems with chiral symmetry (satisfying $\Gamma H^\dag({k})\Gamma^{-1}= -H({k})$~\cite{Kawabata2019a}) is included in the Supplemental Material~\cite{Suppl}.

\begin{figure*}
\includegraphics[width=0.99\textwidth]{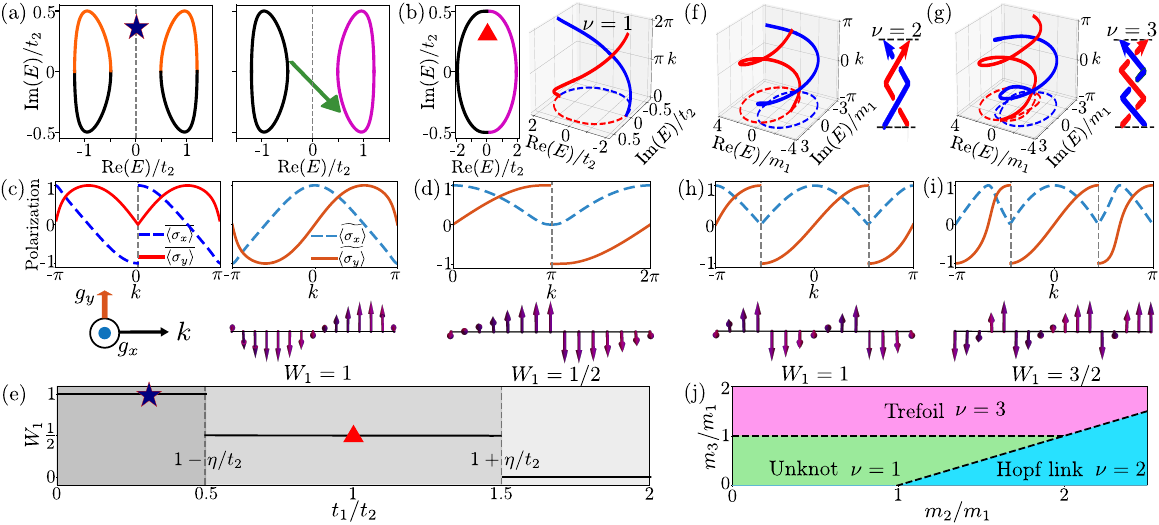}
\caption{Measuring the NH topology of 1D models. (a)-(e) Characterization of the NH SSH model.
Complex-band structures in the real-line-gapped (a) and point-gapped (b) phases are shown, respectively. 
The corresponding dynamical measurements are illustrated in (c) and (d).
In both (a) and (c), the left (right) subfigures represent the measurement before (after) the deformation defined in Eq. (5). 
The deformation maps the measured eigenvectors—whose eigenenergies are colored in orange—in the lower band to their sublattice-symmetric counterparts, as indicated by the green arrow.
Through the mapping, the deformed measurement covers a single band (a) or only half of the band (b), as highlighted in magenta. The dynamical field ${\bm g}(k)=(g_x,g_y)$ (purple arrows) constructed from the deformed LTSTs $\widetilde{\langle\sigma_{x,y}(k)\rangle}$ exhibits the integer-valued winding (c) and the half-integer-valued winding with a $\pi$ jump (d), respectively.
Such measurements can precisely determine the phase diagram (e). Here, we set $\eta=0.5t_2$ and take $t_1=0.3t_2$  (a) or $t_2$ (b). 
(f)-(j) Characterization of the twister Hamiltonian. Complex-band structures in the Hopf link phase (f) and the trefoil phase (g) are shown, respectively, with the corresponding braid diagrams. The dynamical measurements yield nonzero windings of ${\bm g}(k)$ with two (h) or three (i) jumps. The phase
diagram as function of $m_2$ and $m_3$ is plotted (j). Here, we take $(m_2/m_1, m_3/m_1)=(2,0.5)$ in (f) and $(1,1.5)$ in (g). 
}\label{fig1}
\end{figure*}

{\em Dynamical measurement scheme.---}
We start with 1D NH systems with SLS to illustrate our scheme.
The Hamiltonian is generally written as
\begin{align}~\label{H1D}
\begin{split}
H(k)&= H_0(k)+\ui H_1(k), \\
H_0(k)&=\sum_{i=x,y} h_i(k)\sigma_i,\quad H_1(k)=\sum_{i=x,y} h'_i(k)\sigma_i,
\end{split}
\end{align}
for which $S=\sigma_z$. Here $\sigma_{x,y,z}$ are the Pauli matrices and $h'_{x,y}$ can be constant including zero.
The right eigenvectors of $H(k)$ are denoted as $|u_{n=\pm}^{\rm R}(k)\rangle$ with eigenvalues $E_{\pm}(k)=\pm {\varepsilon}(k)$.
Here ${\varepsilon}(k)=[\sum_{i=x,y}(h_i+\ui h'_i)^2]^{-1/2}$ for real-line-gapped $H(k)$.
We investigate the time evolution governed by $H(k)$ starting from a chosen initial fully-polarized state $\ket{\psi_0(k)}$, 
which can be realized experimentally by applying an additional Zeeman field and suddenly changing its
strength from a large value to zero.

To measure the NH topology, we introduce an auxiliary Hermitian matrix $Q(k)$:
\begin{align}~\label{Q_def}
Q(k)=\frac{1}{2}\left[\frac{H(k)}{\varepsilon(k)}+\frac{H^\dag(k)}{\varepsilon^*(k)}\right],
\end{align}
and define the {\em normalized} long-time spin textures (LTSTs)
\begin{equation}~\label{LTST_def}
\overline{\langle\sigma_i(k)\rangle}=\frac{{\langle\sigma_i(k)\rangle}_\infty}{\sqrt{\sum_{j=x,y} \langle\sigma_j(k)\rangle_\infty^2}} ,  \quad  i=x,y,
\end{equation}
where
$\langle\sigma_i\rangle_\infty=\lim_{t\to \infty}{\cal N}_k^{-1}{\rm Tr}(e^{-\ui Ht}\rho_0 e^{\ui H^{\dagger}t}\sigma_i)$ with ${\cal N}_k\equiv{\rm Tr}(e^{-\ui Ht}\rho_0e^{\ui H^{\dagger}t})$ and the density matrix of the initial state $\rho_0=\ket{\psi_0}\bra{\psi_0}$.
After some derivation, we find that (i) $H(k)$ is topologically equivalent to $Q({k})$ in terms of a line (real or imaginary) gap~\cite{Suppl}, and (ii) regardless of the initial state $\rho_0$~\cite{footnote1}, the topology of the Hermitian $Q$ matrix is characterized by the winding number of a dynamical field ${\bm g}(k)=(g_x(k),g_y(k))$:
\begin{align}~\label{W1}
W_1=\frac{1}{2\pi}\int_{\rm BZ} d{k}\left[g_x(k)\partial_{{k}} g_y(k) - g_y(k)\partial_{{k}}g_x(k)\right],
\end{align}
where the integral is over the first Brillouin zone (BZ) and ${\bm g}(k)$ is defined either by $g_i(k)=\overline{\langle\sigma_i(k)\rangle} $ for imaginary line gaps, or by the {\em deformed}  LTSTs as
\begin{align}~\label{gi_def}
g_i(k)=\widetilde{\langle\sigma_i (k)\rangle}= 
\left\{ \begin{array}{ll}
\overline{\langle\sigma_i(k)\rangle} & \mathrm{Im}[\varepsilon({k})]>0\\
-\overline{\langle\sigma_i(k)\rangle} & \mathrm{Im}[\varepsilon({k})]<0
\end{array} \right.
,\,\, i=x,y,
\end{align}
for real line gaps~\cite{Suppl}. Therefore, one can characterize the NH topology of $H(k)$ by measuring $W_1$ from LTSTs.
Although it is introduced to detect the line-gap topology, $W_d$ can, remarkably, also characterize the point-gap topology of $H(k)$, which will be detailed below.
We emphasize that all the LTSTs introduced above are defined only in the basis of right eigenvectors $\{|u_n^{\rm R}(k)\rangle\}$, 
and thus can be directly measured in experiment.

{\em 1D examples.---} We first consider the paradigmatic NH Su-Schrieffer-Heeger (SSH) model~\cite{Lieu2018,Yin2018}, 
which is described by the Hamiltonian~\eqref{H1D} 
with $(h_x,h_y)=(t_1 +t_2\cos k,t_2\sin k)$ and $(h'_x,h'_y)=(0,\eta)$.
Here $t_1$ ($t_2$) describes the tunnelling within (between) unit cells, and $\eta$ represents the asymmetric hopping.
When $t_1<t_2-\eta$ or $t_1>t_2+\eta$ (assuming $t_2>\eta>0$), this system features a real line gap.
An example is shown in Fig.~\ref{fig1}(a). 
The normalized LTSTs $\overline{\langle\sigma_{x,y}(k)\rangle}$ are obtained by investigating the quench dynamics of a fully polarized initial state 
$\rho_0 = (\id-\sigma_z)/2 $~\cite{footnote1} [left subfigure of Fig.~\ref{fig1}(c)]. 
At $k=0$, one can observe a discontinuity and a cusp in $\overline{\langle\sigma_{x,y}(k)\rangle}$, respectively.
This result arises because after long-term dynamics, only the right eigenvectors with ${\rm Im}(E_n)>0$ persist and they belong to different energy bands, as highlighted in 
orange in the left subfigure of Fig.~\ref{fig1}(a).
To accurately characterize the band topology, we note that each right eigenvector $|u_n^{\rm R}\rangle$ with eigenenergy $E_n$ has a sublattice-symmetric counterpart $S|u_n^{\rm R}\rangle$ with $-E_n$, which suggests the deformation defined in Eq.~\eqref{gi_def}~\cite{Suppl}. The deformed LTSTs $\widetilde{\langle\sigma_{x,y}(k)\rangle}$ then capture the correct information of a single band [colored in magenta in the right subfigure of Fig.~\ref{fig1}(a)], and the nonzero winding of the constructed dynamical field ${\bm g}(k)$ characterizes the line-gap topology with $W_1=1$ [Fig.~\ref{fig1}(c)].
An example of characterizing imaginary-line-gapped topological phases is given in~\cite{Suppl}.

One notable finding is that the point-gap topology of this NH SSH model can also be characterized by $W_1=1/2$, indicating the unknotted braiding of complex-energy bands~\cite{Hu2021,WangK2021,Yu2022,ZhangQ2023,WuY2023,Cao2023}.
As illustrated in Fig,~\ref{fig1}(b), when $t_2-\eta<t_1<t_2+\eta$,
the eigenenergy trajectories $E_{\pm}(k)$ braid once to form a single loop in the $({\rm Re}(E),{\rm Im}(E), k)$ space (an unknot~\cite{Hu2021}), resulting in a band structure featuring a point gap. This unknotted braiding implies that our measurement via $\widetilde{\langle\sigma_i (k)\rangle}$ represents a half-period measurement, effectively covering only half of the closed-loop spectrum, as highlighted in magenta in Fig.~\ref{fig1}(b).
Consequently, the dynamical field ${\bm g}(k)$, constructed using the same approach as in the line-gapped case, exhibits incomplete winding characterized by an abrupt $\pi$ jump, yielding $W_1=1/2$ [Fig.~\ref{fig1}(d)].
This half-integer winding number also reflects the vorticity of an EP~\cite{Leykam2017,Shen2018,Yin2018}.

In addition to the unknot discussed above, our dynamical measurement can unveil other types of complex-energy braiding.
For illustration, we consider the twister Hamiltonian that retains the same form as Eq.~\eqref{H1D}:
$H(k)=m_1T_1+m_2T_2+m_3T_3,$ where $T_n=\ket{\uparrow}\bra{\downarrow}+e^{\ui nk}\ket{\downarrow}\bra{\uparrow}$~\cite{Hu2021}.
The phase diagram is depicted in Fig.~\ref{fig1}(j), where
three distinct phases are identified: the unknot, the Hopf link, and the trefoil~\cite{Hu2021}.
Remarkably, we find that all these phases can be detected through our dynamical measurement scheme.
As demonstrated in Figs.~\ref{fig1}(h) and (i), the winding number $W_1$, in conjunction with varying numbers of $\pi$ jumps, can uniquely characterize the complex-energy braiding.
Specifically, when $N$ jumps occur (with $N>0$), the winding of ${\bm g}(k)$ gives the braiding degree
\begin{align}~\label{braid_nu}
|\nu|=N=2|W_1|,
\end{align}
where $\nu$ counts how many times the complex bands braid in the $({\rm Re}(E),{\rm Im}(E), k)$ space~\cite{WangK2021}.

\begin{figure}
\includegraphics[width=0.48\textwidth]{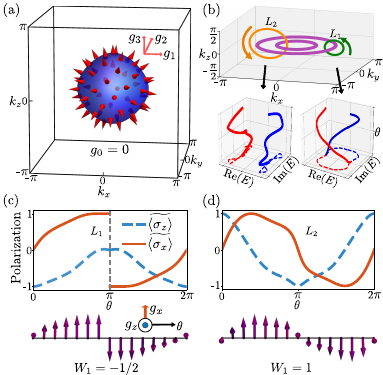}
\caption{Measuring the NH topology of 3D models.
(a) Characterization of the 3D line-gap topology. The dynamical field ${\bm g}({\bm k})$ (arrows) exhibits a hedgehog configuration on the $g_0=0$ surface. Here $m_0=4t_0$, $t_{\rm so}=0.5t_0$, and $\eta=0.2t_0$.
(b)-(d) Characterization of the NH nodal-line semimetal, 
which features two exceptional rings in the BZ, as highlighted in magenta in (b). Along the loop $L_1$ ($L_2$), the vorticity of the enclosed EP (EPs) takes a half-integer (an integer) value, which is revealed by the winding of the dynamical field ${\bm g}(k)=(g_z,g_x)$ constructed from $\widetilde{\langle\sigma_{z,x}(k)\rangle}$ in (c) [(d)].
The evolution of complex eigenvalues is parameterized by $\theta\in[0,2\pi)$.
}\label{fig2}
\end{figure}


{\em Higher dimensions.---} The present scheme can be extended to higher dimensions.
For odd $d$-dimensional ($d$D) line-gapped systems, the Hamiltonian can generally take the form 
\begin{align}~\label{HdD}
\begin{split}
H({\bm k})&= H_0({\bm k})+\ui H_1({\bm k}),\quad H_0({\bm k}) ={\sum}_{i=0}^d h_i({\bm k})\gamma_i, \\
H_1({\bm k})&= h'_{l_1}({\bm k}) \gamma_{l_1}+h'_{l_2}({\bm k}) \gamma_{l_2}+\cdots,\,\, l_i\in\{0,1,\dots,d\}.
\end{split}
\end{align}
Here the $\gamma$ matrices obey the Clifford algebra $\{\gamma_i,\gamma_j\}=2 \delta_{ij}\id$ ($i,j=0,1,\dots,d$) and are
of dimensionality $2^{(d+1)/2}$ for an odd $d$, which is the minimal requirement to open a line gap.
Such a system has the SLS $S=\ui^{(d+1)/2}\prod_{i=0}^{d}\gamma_i$. Similar to 1D systems, 
the topology of $H({\bm k})$ can be dynamically investigated by introducing the  auxiliary $Q$ matrix.
We find that when choosing $\rho_0=(\id-S)/2$ at each ${\bm k}$, the line-gap topology is characterized by a $d$D winding number~\cite{Suppl}
\begin{align}~\label{Wd}
W_d=\frac{[(d-1)/2]!}{2\pi^{(d+1)/2}d!}\int_{{\rm BZ}^d}{\bm g}({\rm d}{\bm g})^{d},
\end{align}
where the dynamical field ${\bm g}=(g_0,g_1,\dots,g_d)$ with $g_i({\bm k})=\overline{\langle\gamma_i ({\bm k})\rangle}$ [or $\widetilde{\langle\gamma_i ({\bm k})\rangle}$ as given in Eq.~\eqref{gi_def}] for imaginary (real) line gaps, ${\bm g}(\ud{\bm g})^{d}\equiv\epsilon^{i_{1}i_{2}\cdots i_{d+1}}g_{i_{1}}\ud g_{i_{2}}\wedge\cdots\wedge\ud g_{i_{d+1}}$
with $\epsilon^{i_{1}i_{2}\cdots i_{d+1}}$ being the fully antisymmetric tensor and $i_{1,2,\dots,d+1}\in\{0,1,2,\dots,d\}$,
and `$\ud$' denotes the exterior derivative.

We illustrate it with three-dimensional (3D) NH topological insulators.
The Hamiltonian takes the form of Eq.~\eqref{HdD} with $h_0= m_0 -2t_0\sum_{i=1}^{3} \cos k_{r_i}$, $h_{i>0}=2t_{\rm so}\sin k_{r_i}$, and $h_2'=\eta$.
Here we denote $(r_1,r_2,r_3) \equiv (x,y,z)$ and take $\gamma_0 = \sigma_z \otimes \tau_x$, $\gamma_1 = \sigma_x \otimes\id$, $\gamma_2 = \sigma_y \otimes\id$, and $\gamma_3 = \sigma_z \otimes \tau_y$, where $\sigma_i$ and
$\tau_i$ are both Pauli matrices.
The Hermitian part $H_0({\bm k})$ has been simulated in solid-state spin
systems~\cite{Ji2020,Xin2020}.
The line-gap topology can be unveiled through the winding of the dynamical field ${\bm g}({\bm k})=(g_0,g_1, g_2, g_3)$, with $g_i=\widetilde{\langle\gamma_i\rangle}$, across the entire BZ.
One can find that on each constant-$g_0$ surface---such as the $g_0=0$ surface shown in Fig.~\ref{fig2}(a)---${\bm g}({\bm k})$ displays a hedgehog configuration, indicating a nonzero winding $W_3=1$~\cite{Suppl}.

Besides the gapped cases, the proposed scheme can also be applied to topological semimetals. 
Here, we consider the 3D NH nodal-line semimetals~\cite{Wang2019,Yang2019}:
$H({\bm k})=[m+t(\cos k_x+\cos k_y +\cos k_z)]\sigma_z+(t\sin k_z +\ui\eta)\sigma_x$.
When $m=-2t$ and $\eta=-0.5t$, the system features two exceptional rings in the $k_z=0$ plane [see Fig.~\ref{fig2}(b)].
As each ring comprises a collection of EPs, the vorticity of an EP can be used to characterize these exceptional rings~\cite{Wang2019}.
For a closed loop encircling one ring (two rings), our dynamical measurement can 
give exactly a half-integer (integer) vorticity [Fig.~\ref{fig2}(c) and (d)].

\begin{figure}
\includegraphics[width=0.485\textwidth]{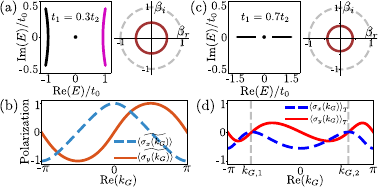}
\caption{Measuring non-Bloch topological invariants. (a) The OBC spectrum (left) and GBZ (right) of the NH SSH model with $t_1=0.3t_2$. 
Here $\beta_r$ ($\beta_i$) denotes the real (imaginary) part of $\beta$.
The dashed unit circle represents the BZ.
(b) The measurement of  $\widetilde{\langle\sigma_{x,y}(k_G)\rangle}$ with $\rho_0=\ket{x_0}\bra{x_0}\otimes(\id-\sigma_z)/2$~\cite{footnote3}
 yields the winding number $W_1=1$, which characterizes the zero modes in (a). Here, $x_0$ represents the lattice center site.
(c) The real-energy spectrum (left) and GBZ (right) for $t_1=0.7t_2$. 
(d) Time-averaged spin textures $\langle\sigma_{x,y}(k_G)\rangle_T$ with $\rho_0 =\ket{x_0}\bra{x_0}\otimes(\id-\sigma_x)/2$. 
The slopes of $\langle\sigma_{y}(k_G)\rangle_T$ at $k_{G,1}$ and $k_{G,2}$ determine $C_0=1$, characterizing the zero modes in (c).
}\label{fig3}
\end{figure}

{\em Generalized Brillouin zone.---} The above discussion only considers periodic boundary conditions (PBCs). 
Here we generalize the present scheme to open boundary conditions (OBCs).
As proposed in Ref.~\cite{Yao2018a}, we will deal with it in the GBZ.
Since only the construction of 1D GBZ is well established~\cite{Yao2018a,Yokomizo2019}, we will restrict our discussion to 1D cases here. 

In 1D GBZ theory~\cite{Yao2018a}, the non-Bloch Hamiltonian $H(k_{G})$, corresponding to $H(k)$ in Eq.~\eqref{H1D}, is obtained by the replacement $e^{\ui k}\to\beta\equiv e^{\ui k_{G}}$, where $k_{G}$ is complex.
The auxiliary $Q$ matrix, which is topologically equivalent to $H(k_G)$, can be accordingly introduced as in Eq.~\eqref{Q_def}~\cite{Suppl}.
For an initial state $\rho_0(x)=\ket{\psi_0(x)}\bra{\psi_0(x)}$ in real space, we introduce a projection operator $P_{k_G}\equiv|\beta^{\rm R}(k_G)\rangle\langle \beta^{\rm L}(k_G)|$ 
to project the time evolution $\rho(x,t)=e^{-\ui H t}\rho_0(x)e^{\ui H^\dag t}$ onto the GBZ: $\rho(k_G,t)=P_{k_G}\rho(x,t)P_{k_G}^\dag$~\cite{LiT2021},
where $|\beta^{\rm R}(k_G)\rangle=L^{-1/2}\sum_i e^{ik_G x_i}\ket{x_i}$ and $|\beta^{\rm L}(k_G)\rangle=L^{-1/2}\sum_i e^{ik_G^* x_i}\ket{x_i}$, with $L$ the lattice length, which satisfy $\langle \beta^{\rm L}(k_G)|\beta^{\rm R}(k'_G)\rangle=\delta_{k_G,k'_G}$ and $\sum_{k_G}|\beta^{\rm R}(k_G)\rangle\langle \beta^{\rm L}(k_G)|=\id$~\cite{footnote2}. This gives the normalized LTSTs $\overline{\langle\sigma_{x,y}(k_G)\rangle}$, 
as defined in Eq.~\eqref{LTST_def}, with the replacement $e^{-\ui H t}\rho_0(k)e^{\ui H^\dag t}\to\rho(k_G,t)$.

The dynamical measurement in the GBZ is divided into two cases: 
(i) When the OBC spectrum is complex, the procedure follows exactly as outlined in Eqs.~\eqref{W1} and \eqref{gi_def}.
An illustrative example is provided in Figs.~\ref{fig3}(a) and (b), where we examine the 1D NH SSH model under OBCs, using the same parameters as those in Fig~\ref{fig1}(a).
(ii) In contrast, when the spectrum is entirely real,
the situation becomes analogous to the characterization of Hermitian systems~\cite{Zhang2018}.
Rather than using LTSTs, we introduce the {\it time-averaged} spin textures:
\begin{align}
\langle\sigma_i(k_G)\rangle_T=\lim_{T\to \infty}\int_0^T\frac{1}{{\cal N}_{k_G}}{\rm Tr}\left[\rho(k_G,t)\sigma_i\right] dt,\,\,\, i=x,y,
\end{align}
where ${\cal N}_{k_G}\equiv{\rm Tr}\left[\rho(k_G,t)\right]$. We prove that when choosing the initial spin state to be $(\id-\sigma_x)/2$~\cite{footnote3},
the open-boundary topology is characterized by a zero-dimensional topological invariant~\cite{Suppl}
\begin{align}
C_0=\frac{1}{2}\sum_{j}\mathrm{sgn}\left[\partial_{k_G} {\langle\sigma_{y}(k_{G,j})\rangle}_T\right],
\end{align}
where $k_{G,j}$ denote the paired points at which ${\langle\sigma_{y}(k_{G,j})\rangle}_T=0$ and $|{\langle\sigma_{x}(k_{G,j})\rangle}_T|$ take local minima.
An example is given in Fig.~\ref{fig3}(c) and (d), and more details can be found in \cite{Suppl}.
{\color{red} }

\begin{figure}
\includegraphics[width=0.49\textwidth]{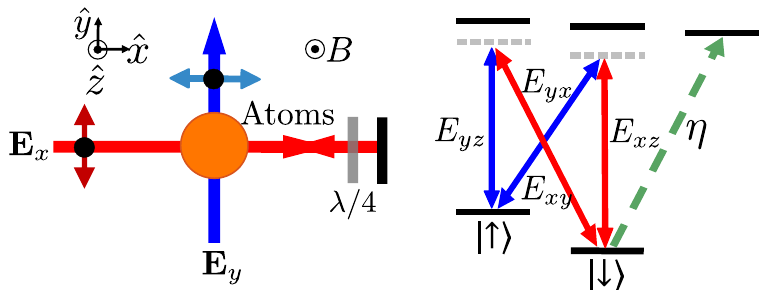}
\caption{Experimental setup and laser couplings. 
A standing wave ${\bm E}_{x}$ (red) and a plane wave ${\bm E}_{y}$ (blue) form a double-$\Lambda$-type configuration (right panel)
to generate both lattice and Raman  potentials. 
A near-resonant beam (green) is added to induce a spin-dependent atom loss.
}\label{fig4}
\end{figure}

Before proceeding, several important points should be emphasized:
(i) As described above, a well-chosen initial state can significantly simplify the measurement, with the selection highly depending on the system’s dimensionality and symmetry. A summary can be found in \cite{Suppl}.
(ii) Although we assume the initial state to be completely
polarized for simplicity, the proposed scheme is also applicable in the case of a “shallow quench”, where the initial state is not fully polarized but remains within the same trivial regime as the fully-polarized state. This robustness has been both theoretically analyzed 
and experimentally demonstrated in the dynamical characterization of Hermitian systems~\cite{Zhang2019b,Yi2019}.
(iii) For OBCs, the extension to higher-dimensional GBZ is straightforward. 
Once the GBZ is determined through a theoretical framework, such as the theory established in Refs.~\cite{Wang2024,Xiong2024}, one can
project the measured spin textures onto the corresponding GBZ, thereby enabling the
characterization of the non-Bloch topology.

{\em Experimental realization.---} Based on optical Raman lattices~\cite{LiuXJ2013,Wang2018,Sun2018,Song2018,LiangMC2023}, we propose a highly feasible experimental setup to realize and detect 1D NH topological phases in ultracold atoms.
As sketched in Fig.~\ref{fig4}, a standing-wave beam  ${\bm E}_{x}$ (green) with  $x,z$ polarization
and a plane wave ${\bm E}_{y}$ (orange) with $y,z$ polarization are applied to form a double-$\Lambda$-type configuration,
which generates a spin-independent lattice $V(x)$ and two Raman coupling potentials ${\cal M}_{x,y}(x)$ simultaneously.
A $\lambda/4$ wave plate is placed before the mirror to induces an additional phase shift to the $z$ component light, which results in
${\bm E}_x=\hat{y}E_{xy}\cos(k_0x)+\ui\hat{z}E_{xz}\sin(k_0 x)$. Here $k_0=2\pi/\lambda$ with $\lambda$ being the wavelength.
Thus, the Raman potential ${\cal M}_x$ (${\cal M}_y$), generated by the components $E_{xy}$ and $E_{yz}$ ($E_{xz}$ and $E_{yx}$), takes the form
${\cal M}_{x}(x)=M_{0x}\cos(k_0x)$ (${\cal M}_{y}(x)=M_{0y}\sin(k_0x)$), with $M_{0x}\propto E_{xy}E_{yz}/\Delta$ ($M_{0y}\propto E_{xz}E_{yx}/\Delta$), where $\Delta$ denotes the one-photon detuning.
The lattice potential reads $V(x)=V_0\cos^2(k_0x)$, with $V_0\propto(E_{xz}^2-E_{xy}^2)/\Delta$.
In addition, a single near-resonant beam is added to induce a spin-selective atom loss $\eta$~\cite{Ren2022,Zhao2023}.
The total Hamiltonian then reads~\cite{Suppl}
\begin{align}~\label{Ham_exp}
\begin{split}
H=&\left[\frac{k_x^2}{2m}+V_0\cos^2(k_0x)\right]\otimes\id+M_{0x}\cos(k_0x)\sigma_x \\
&+M_{0y}\sin(k_0x)\sigma_y+m_z\sigma_z-\ui\eta\ket{\downarrow}\bra{\downarrow},
\end{split}
\end{align}
where $m$ is the mass of an atom and $m_z$ measures the two-photon detuning.
Note that the Raman ${\cal M}_x$ (${\cal M}_y$) and lattice potentials satisfy a relative antisymmetry (symmetry) with respect to each lattice site,
ensuring that ${\cal M}_x$ (${\cal M}_y$) only induces spin-flip hopping (on-site spin flipping).
The Bloch Hamiltonian in the tight-binding limit then takes
$H(q)=(m_z-2t_0\cos qa){\sigma}_z+(m_y+2t_{\rm so}\sin qa){\sigma}_y+\ui\eta/2(\sigma_z-\id)$,
which describes a NH SSH-like model that hosts a point-gap topological phase~\cite{Suppl}.
Here $a$ is the lattice constant, $q$ is the quasimomentum, and $t_0$ ($t_{\rm so}$) represents the spin-conserved (spin-flip) hopping. 
The model parameters can be tuned independently.

The proposed dynamical measurement scheme can be implemented within this cold-atom system.
Numerical results based on the Hamiltonian~\eqref{Ham_exp} are detailed in the Supplementary Material~\cite{Suppl}, 
where the initial state is a wave packet localized at the center of the lattice, 
and measurements are projected onto the BZ (GBZ) to characterize the topology under PBCs (OBCs).
It is crucial to note that while the LTSTs are defined in the infinite-time limit [Eq.~\eqref{LTST_def}], the characterization can be accurate enough after a finite time $\tau^*\gg1/\eta$.
Importantly, the time scale $\tau^*$ can often be significantly shorter than the lifetime of the cold-atom system.
Taking $^{40}$K atoms as an example, one can choose $\ket{\uparrow}=|F=7/2, m_F=7/2\rangle$, $\ket{\downarrow}=|9/2, 9/2\rangle$, and set $\lambda=$768nm.
The time scale for obtaining $\widetilde{\langle\sigma_{y,z}\rangle}$ is $\tau^*=120/E_r$ with $E_r=k_0^2/(2m)$~\cite{Suppl}, while the lifetime for $^{40}$K atoms is about
$\tau\sim$1300$/E_r$~\cite{Wang2018}. The requirement $\tau^*\ll\tau$ is indeed satisfied, ensuring that there is ample time available for implementing the dynamical measurement.
A detailed illustration based on $^{40}$K atoms is provided in \cite{Suppl}, while the proposed scheme 
is also suitable for other alkali-metal atoms, such as $^{87}$Rb bosons, as well as alkaline-earth(-like) atoms, such as $^{173}$Yb and $^{87}$Sr.



{\em Conclusion.---} We have proposed a generic and unified scheme to directly measure line-gap winding numbers, point-gap braiding degrees, and non-Bloch topological invariants in odd-dimensional systems.
The experimental setup we've designed not only demonstrates the practicality of detecting 1D NH topological phases but also highlights the accessibility of these measurements with existing technologies.
Despite only odd-dimensional systems being considered, the proposed scheme can certainly be generalized to NH systems in even dimensions,
which will be our next study.

We acknowledge insightful discussions with Haiping Hu and Xiong-Jun Liu.
This work was supported by the National Natural Science Foundation of China (Grants No. 12204187),  
the Innovation Program for Quantum Science and Technology (Grant No. 2021ZD0302000), 
and the startup grant of Huazhong University of Science and Technology (Grant No. 3034012114).

\setcounter{equation}{0} \setcounter{figure}{0}
\setcounter{table}{0} 
\renewcommand{\theparagraph}{\bf}
\renewcommand{\thefigure}{S\arabic{figure}}
\renewcommand{\theequation}{S\arabic{equation}}
\renewcommand{\thetable}{S\arabic{table}}

\onecolumngrid
\flushbottom
\newpage

\section*{\large SUPPLEMENTAL MATERIAL}

\section{I.\quad Non-Hermitian systems with sublattice symmetry}

For odd-dimensional non-Hermitian (NH) systems exhibiting sublattice symmetry: $SH({\bm k})S^{-1}= -H({\bm k})$~\cite{Kawabata2019_S}, the Hamiltonian can generally be written as
\begin{align}~\label{General_H_SLS}
	\begin{split}
		H({\bm k})&= H_0({\bm k})+\ui H_1({\bm k}),\quad H_0({\bm k}) ={\sum}_{i} h_i({\bm k})\gamma_i, \quad H_1(\bm{k})=\sum_{i} h'_i(\bm{k})\gamma_i,
	\end{split}
\end{align}
where $h'_{i}$ can be constant including zero, and the $\gamma$ matrices obey the Clifford algebra $\{\gamma_i,\gamma_j\}=2 \delta_{ij}\id$ and are of dimensionality $2^{(d+1)/2}$ for odd $d$-dimensional ($d$D) line-gapped insulators. The right eigenvectors of NH Hamiltonian \eqref{General_H_SLS} are denoted as $\ket{u_{n^{(\pm)}}^R(\bm{k})}$ with eigenvalues $E_{n^{(\pm)}}(\bm{k})=\pm \varepsilon (\bm{k}) $. Here $ {\varepsilon}(\bm{k})=\sqrt{\sum_{i} [h_i(\bm{k})+\ui h'_i(\bm{k})]^2}$ for real-line-gapped phases.

\subsection{A.\quad Definition of the $Q$ matrix and proof of the topological equivalence}\label{SecIA}

For the sake of simplicity, we focus on one-dimensional (1D) systems in this subsection. Nonetheless, both the results and the proof can be easily extended to higher-dimensional systems. For a 1D NH Hamiltonian $H(k)$, the right and left eigenvectors are defined by
\begin{align}
H(k)|u_{\pm}^{\rm R}\rangle=\pm\varepsilon(k)|u_{\pm}^{\rm R}\rangle,\quad H^\dag(k)|u_{\pm}^{\rm L}\rangle=\pm\varepsilon^*(k)|u_{\pm}^{\rm L}\rangle.
\end{align}
When $\varepsilon(k)\neq0$, we introduce the $Q$ matrix as
\begin{align}
\begin{split}
&Q(k)=\id-P_{-}^{\rm R}(k)-P_{-}^{\rm L}(k), \\
&P_{\pm}^{\rm R}\equiv|u_{\pm}^{\rm R}\rangle\langle u_{\pm}^{\rm L}|,\quad P_{\pm}^{\rm L}\equiv|u_{\pm}^{\rm L}\rangle\langle u_{\pm}^{\rm R}|.
\end{split}
\end{align}
Since $P_{+}^{\rm R}+P_{-}^{\rm R}=\id$ and $H/\varepsilon=P_{+}^{\rm R}-P_{-}^{\rm R}$, we have $2P_{-}^{\rm R}=\id-H/\varepsilon$. 
Similarly, one can derive $2P_{-}^{\rm L}=\id-H^\dag/\varepsilon^*$. The $Q$ matrix is then rewritten as
\begin{align}~\label{Qk1}
Q(k)=\frac{1}{2}\left[\frac{H(k)}{\varepsilon(k)}+\frac{H^\dag(k)}{\varepsilon^*(k)}\right].
\end{align}

We then prove the topological equivalence between $H$  and $Q$.
We define 
\begin{align}
{\cal H}(s)\equiv  s\frac{H}{\varepsilon}+(1-s)Q=\frac{1+s}{2}\frac{H}{\varepsilon}+\frac{1-s}{2}\frac{H^\dag}{\varepsilon^*},
\end{align}
where $s\in[0,1]$. One can derive that
\begin{align}\label{Hsproof}
{\cal H}^2&=\frac{(1+s)^2}{4}\left(\frac{H}{\varepsilon}\right)^2+\frac{(1-s)^2}{4}\left(\frac{H^\dag}{\varepsilon^*}\right)^2+\frac{1-s^2}{4}\frac{HH^\dag+H^\dag H}{|\varepsilon|^2}\nonumber\\
&=\frac{(1+s)^2}{2}+\frac{1-s^2}{2}\frac{[{\rm Re}(H)]^2+[{\rm Im}(H)]^2}{|\varepsilon|^2} \nonumber\\
&\geq \frac{(1+s)^2}{2}.
\end{align}
We then have ${\cal H}^2>0$, i.e., ${\cal H}(s)$ is always gapped for $0\leq s\leq 1$, which means $H/\varepsilon$ and $Q$ are topologically equivalent.
Therefore, $H$ is topologically equivalent to $Q$ in terms of line gaps.

\subsection{B.\quad Dynamical measurement scheme}

We define the normalized long-time spin textures (LTSTs) as
\begin{equation}~\label{LTST_def}
	\overline{\langle\gamma_i(\bm{k})\rangle}=\frac{{\langle\gamma_i(\bm{k})\rangle}_\infty}{\sqrt{\sum_j \langle\gamma_j(\bm{k})\rangle_\infty^2}},
\end{equation}
where
$\langle\gamma_i\rangle_\infty=\lim_{t\to \infty}{\cal N}_{\bm{k}}^{-1}{\rm Tr}(e^{-\ui Ht}\rho_0 e^{\ui H^{\dagger}t}\gamma_i)$, with the normalization factor ${\cal N}_{\bm{k}}\equiv{\rm Tr}(e^{-\ui Ht}\rho_0e^{\ui H^{\dagger}t})$ and a chosen initial state $\rho_0$. Since $e^{-\ui Ht} = \cos(\varepsilon t)-\ui\sin(\varepsilon t)H/\varepsilon$, one can prove that
after long-term dynamics, the LTSTs are given by
\begin{align}~\label{LTST0}
	\langle\gamma_i(\bm{k})\rangle_\infty =
	\left\{ \begin{array}{ll}
		{\cal C}_1^{-1}{\rm Tr}\left[\rho_0 \gamma_i(1+ \frac{H}{\varepsilon} - \frac{H^\dagger}{\varepsilon^*}-\frac{H^\dagger H}{\varepsilon^* \varepsilon} + 2\frac{h_i-\ui h'_{i}}{\varepsilon^*}\gamma_i+2\frac{h_i-\ui h'_{i}}{\varepsilon^* \varepsilon}\gamma_i H)\right] &\quad \mathrm{Im}[\varepsilon(\bm{k})]>0\\
		{\cal C}_2^{-1}{\rm Tr}\left[\rho_0 \gamma_i(1- \frac{H}{\varepsilon} + \frac{H^\dagger}{\varepsilon^*}-\frac{H^\dagger H}{\varepsilon^* \varepsilon} - 2\frac{h_i-\ui h'_{i}}{\varepsilon^*}\gamma_i+2\frac{h_i-\ui h'_{i}}{\varepsilon^* \varepsilon}\gamma_i H)\right] &\quad \mathrm{Im}[\varepsilon({\bm{k}})]<0
	\end{array} \right. ,
\end{align}  
with ${\cal C}_1={\rm Tr}\left[\rho_0(1+ \frac{H}{\varepsilon} + \frac{H^\dagger}{\varepsilon^*}+\frac{H^\dagger H}{\varepsilon^* \varepsilon})\right]$ and ${\cal C}_2={\rm Tr}\left[\rho_0(1- \frac{H}{\varepsilon} - \frac{H^\dagger}{\varepsilon^*}+\frac{H^\dagger H}{\varepsilon^* \varepsilon})\right]$. 
When the initial state is prepared as $\rho_0=(\id-S)/2$ at each ${\bm k}$, we have
\begin{align}~\label{LTST1}
	\langle\gamma_i(\bm{k})\rangle_\infty =
	\left\{ \begin{array}{ll}
		{\cal C}_1^{-1}\left[\frac{h_i+\ui h'_i}{\varepsilon} + \frac{h_i-\ui h'_i}{\varepsilon^*} -\frac{1}{2}\sum_{j\neq i}{\rm Tr}(S\gamma_i \gamma_j) (\frac{h_j+\ui h'_j}{\varepsilon} - \frac{h_j-\ui h'_j}{\varepsilon^*})\right] & \quad \mathrm{Im}[\varepsilon(\bm{k})]>0\\
		-{\cal C}_2^{-1}\left[\frac{h_i+\ui h'_i}{\varepsilon} + \frac{h_i-\ui h'_i}{\varepsilon^*} -\frac{1}{2}\sum_{j\neq i}{\rm Tr}(S\gamma_i \gamma_j) (\frac{h_j+\ui h'_j}{\varepsilon} - \frac{h_j-\ui h'_j}{\varepsilon^*})\right] & \quad \mathrm{Im}[\varepsilon({\bm{k}})]<0
	\end{array} \right.,
\end{align}  
which yields the normalized LTSTs
\begin{align}~\label{Results}
	\overline{\langle\gamma_i(\bm{k})\rangle} =
	\left\{ \begin{array}{ll}
		\left(\frac{h_i+ih'_i}{\varepsilon} + \frac{h_i-ih'_i}{\varepsilon^*}\right)/\left[\sum_j \left(\frac{h_j+ih'_j}{\varepsilon} + \frac{h_j-ih'_j}{\varepsilon^*} \right)^2\right]^{1/2} & \mathrm{Im}[\varepsilon(\bm{k})]>0\\
		-\left(\frac{h_i+ih'_i}{\varepsilon} + \frac{h_i-ih'_i}{\varepsilon^*}\right)/\left[\sum_j \left(\frac{h_j+ih'_j}{\varepsilon} + \frac{h_j-ih'_j}{\varepsilon^*} \right)^2\right]^{1/2} & \mathrm{Im}[\varepsilon(\bm{k})]<0
	\end{array} \right. .
\end{align}

Note that, while for $d$D NH systems it is necessary to choose $\rho_0 = (\id-S)/2$, in the case of 1D NH systems, the results of $\langle\sigma_i(k)\rangle_\infty$ are independent of the initial state $\rho_0$. To see it, one can project $\rho_0$ onto the eigenstates of the NH Hamiltonian $\rho_0 = \sum_{nm} c_{nm} |u_{n}^R(\bm{k})\rangle\langle u_m^R(\bm{k})|$ with $c_{nm} = \langle u_{n}^L(\bm{k})|\rho_0|u_m^{L}(\bm{k})\rangle$, and $\rho(T) = \sum_{nm} c_{nm} e^{-i(E_{n}-E_{m}^*)T} |u_{n}^R(\bm{k})\rangle\langle u_m^R(\bm{k})|$ after the evolution time $T$. Since $E_{n^{(\pm)}}= \pm \varepsilon(\bm{k})$, one has
\begin{equation}~\label{rhot}
	\begin{aligned}
		\rho(T) &= \sum_{n^{(+)},m^{(+)}} c_{nm} e^{2{\rm{Im}}(\varepsilon T)} \ket{u_{n^{(+)}}^R(\bm{k})}\bra{u_{m^{(+)}}^R(\bm{k})}+\sum_{n^{(-)},m^{(-)}} c_{nm} e^{-2{\rm{Im}}(\varepsilon T)} \ket{u_{n^{(-)}}^R(\bm{k})}\bra{u_{m^{(-)}}^R(\bm{k})}\\
		&+\sum_{n^{(+)},m^{(-)}} c_{nm} e^{-\ui 2{\rm{Re}}(\varepsilon T)} \ket{u_{n^{(+)}}^R(\bm{k})}\bra{u_{m^{(-)}}^R(\bm{k})}+\sum_{n^{(-)},m^{(+)}} c_{nm} e^{\ui 2{\rm{Re}}(\varepsilon T)} \ket{u_{n^{(-)}}^R(\bm{k})}\bra{u_{m^{(+)}}^R(\bm{k})}.
	\end{aligned}
\end{equation}
In the infinite-time limit, only the terms with positive imaginary energies persist in the time evolution, yielding
\begin{align}~\label{rhoT}
	\rho(T\to\infty) =
	\left\{ \begin{array}{ll}
		\sum_{n^{(+)},m^{(+)}} c_{nm} e^{2{\rm{Im}}(\varepsilon T)}  \ket{u_{n^{(+)}}^R(\bm{k})}\bra{u_{m^{(+)}}^R(\bm{k})}&\quad \mathrm{Im}[\varepsilon(\bm{k})]>0\\
		\sum_{n^{(-)},m^{(-)}} c_{nm} e^{-2{\rm{Im}}(\varepsilon T)} \ket{u_{n^{(-)}}^R(\bm{k})}\bra{u_{m^{(-)}}^R(\bm{k})} &\quad \mathrm{Im}[\varepsilon(\bm{k})]<0
	\end{array} \right.,
\end{align}
where the coefficients $c_{nm}$ depend on $\rho_0$. However, in one dimension, there are only two complex-energy bands and the result simplifies to
\begin{align}
	\rho(T\to\infty) =
	\left\{ \begin{array}{ll}
		 \ket{u_{+}^R(\bm{k})}\bra{u_{+}^R(\bm{k})}&\quad \mathrm{Im}[\varepsilon(\bm{k})]>0\\
		\ket{u_{-}^R(\bm{k})}\bra{u_{-}^R(\bm{k})} &\quad \mathrm{Im}[\varepsilon(\bm{k})]<0
	\end{array} \right.,
\end{align}
which is independent of the initial state $\rho_0$.

As introduced in the above subsection, the $Q$ matrix is given by
\begin{align}~\label{Q_Gen}
	Q(\bm{k})=\frac{1}{2}\left[\frac{H(\bm{k})}{\varepsilon(\bm{k})}+\frac{H^\dag(\bm{k})}{\varepsilon^*(\bm{k})}\right].
\end{align}
One can write
\begin{align}
Q({\bm k})=\sum_iQ_i({\bm k})\gamma_i,\,\, {\rm where}\,\, Q_i(\bm{k}) = \frac{1}{2}\left(\frac{h_i+ih'_i}{\varepsilon}+\frac{h_i-ih'_i}{\varepsilon^*} \right),
\end{align}
and easily prove that
\begin{align}~\label{ResultsGen}
	\overline{\langle\gamma_i(\bm{k})\rangle} =
	\left\{ \begin{array}{ll}
		\overline{Q_i(\bm{k})} & \mathrm{Im}[\varepsilon(\bm{k})]>0\\
		-\overline{Q_i(\bm{k})} & \mathrm{Im}[\varepsilon(\bm{k})]<0
	\end{array} \right. ,
\end{align}
where $\overline{Q_i(\bm{k})}={Q_i(\bm{k})}/\sqrt{\sum_j Q_j(\bm{k})^2}$.
This result forms the foundation of our dynamical measurement scheme.
We accordingly introduce the dynamical field $\bm{g}(\bm{k})=(g_0,g_1,\dots,g_d)$ to characterize the topology of the $Q$ matrix, where $g_i(\bm{k})$ is defined either by $g_i(\bm{k})=\overline{\langle\gamma_i(\bm{k})\rangle} $ for imaginary line gaps, or by the deformed LTSTs as
\begin{align}~\label{gi_defGen}
	g_i(\bm{k})=\widetilde{\langle\gamma_i (\bm{k})\rangle}= 
	\left\{ \begin{array}{ll}
		\overline{\langle\gamma_i(\bm{k})\rangle} & \mathrm{Im}[\varepsilon(\bm{k})]>0\\
		-\overline{\langle\gamma_i(\bm{k})\rangle} & \mathrm{Im}[\varepsilon(\bm{k})]<0
	\end{array} \right.
\end{align}
for real line gaps. From Eqs.~\eqref{ResultsGen} and \eqref{gi_defGen}, one can derive $g_i(\bm{k}) =\overline{Q_i(\bm{k})}$. 
Due to the topological equivalence between $H$ and $Q$ (see Sec.~\ref{SecIA}), 
the $d$D winding number $W_d$ of the dynamical field $\bm{g}(\bm{k})$, as defined in Eq.~(8) in the main text, characterizes the line-gap topology of $H(\bm{k})$.

\subsection{C.\quad1D and 3D examples}\label{SecIB}

\begin{figure*}
	\includegraphics[width=0.99\textwidth]{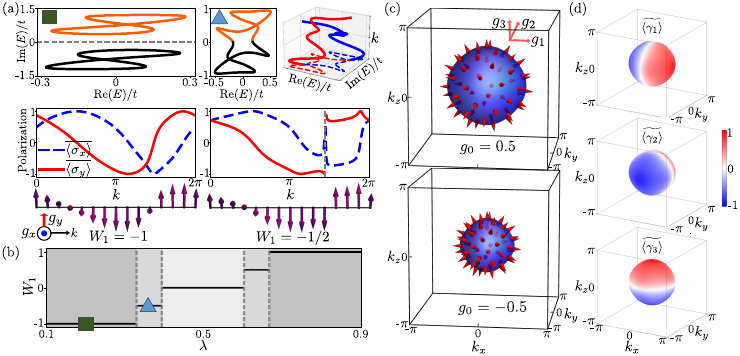}
	\caption{Measuring the topology of NH systems with sublattice symmetry.
(a)-(b) 1D NH model.
Two examples, represented by square and triangle markers, are presented in (a). They correspond to imaginary-line-gapped ($\lambda=0.2$) and point-gapped ($\lambda=0.36$) topological phases, respectively. 
Left panel: The dynamical measurement covers a whole band (orange) and the dynamical field $\bm{g}(k)=(g_x(k),g_y(k))$ (purple arrows), constructed from $\overline{\langle \sigma_{x,y}\rangle}$, exhibits a nonzero winding number $W_1=-1$.
Right panel: The dynamical measurement covers only half of the closed-loop spectrum and the dynamical field $\bm{g}(k)$ displays incomplete winding with an abrupt $\pi$ jump, yielding  $W_1=-1/2$.
The phase diagram as function of $\lambda$ is plotted in (b). Here, $t_1=0.5t$, $t_2=t$, and $m=0.5t$. 
(c)-(d) 3D NH topological insulator. The dynamical field ${\bm g}({\bm k})$ (arrows) exhibits a hedgehog configuration on the $g_0=\pm 0.5$ surfaces (c). 
The deformed LTSTs $\widetilde{\langle\gamma_i ({\bm k})\rangle}$ ($i=1,2,3$) that are used to construct the dynamical field on the $g_0=0.5$ surface are displayed in (d). 
Here, $m_0=4t_0$, $t_{\rm so}=0.5t_0$, and $\eta=0.2t_0$.
	}\label{figS1}
\end{figure*}

We consider a 1D NH Hamiltonian with the Hamiltonian $H(\lambda,k)=\lambda H_0(k) +\ui (1-\lambda)H_1(k)$ controlled by a parameter $\lambda$ ($0\leq\lambda\leq1$), where
\begin{align}~\label{Imag-Line gap}
H_0(k) = (t_1 + t_2 \cos k) \sigma_x + t_2 \sin k \sigma_y,\quad
H_1(k) = (m + t \sin k) \sigma_x + t \cos k \sigma_y.
\end{align}
Here $H_0(k)$ is the paradigmatic Su-Schrieffer-Heeger (SSH) model and more notably, $H_1(k)$ is also a SSH-like model, which is topologically nontrivial when $m<t$. 
The parameter $\lambda$ can be continuously adjusted to transition the system from the real-line-gapped topological phase, dominated by $H_0(k)$ ($\lambda=1$), to the imaginary-line-gapped topological phase, dominated by $H_1(k)$ ($\lambda=0$). In the intermediate region, the complex-energy spectrum can display a point gap, resulting in point-gapped topological phases. The phase diagram is depicted in Fig.\ref{figS1}(b), where the line-gaped (point-gapped) topological phases are characterized by $W_1=\pm1$ ($W_1=\pm1/2$). 

An example of imaginary-line-gapped (point-gapped) topological phase is presented in the left (right) panel of Fig.~\ref{figS1}(a), corresponding to $\lambda=0.2$ ($\lambda=0.36$).
In the imaginary-line-gapped phase, the dynamical field $\bm{g}(k)$ constructed directly from $\overline{\langle\sigma_{x,y}(k)\rangle}$ winds once across the Brillouin zone (BZ), resulting in $W_1=-1$. In contrast, in the point-gapped phase, the eigenenergy trajectories $E_{\pm}(k)$ braid once to form a single loop in the (${\rm{Re}}(E),{\rm{Im}}(E),k$) space. The dynamical field $\bm{g}(k)$, constructed from $\overline{\langle\sigma_{x,y}(k)\rangle}$ using the same method as in the imaginary-line-gapped case, displays a nonzero winding with an abrupt $\pi$ jump, yielding $W_1 = -1/2$. When $0.6<\lambda<0.67$ and $0.67<\lambda<1$, the system exhibits a point-gapped phase with $W_1=1/2$ and a real-line-gapped topological phase with $W_1=1$, respectively, which are similar to those phases of the NH SSH model discussed in the main text.

We now consider the 3D NH topological insulator discussed in the main text and provide additional details.  
Similar to the Hermitian limit, as long as the NH system remains line-gapped, the Hamiltonian $H(\mathbf{k})$ can support 3D NH topological phases, encompassing three regions: region I, where $2t_0<m_0<6t_0$ with winding number $W_3=1$; region II, where $-2t_0<m_0<2t_0$ with winding number $W_3=-2$; and region III, where $-6t_0<m_0<-2t_0$ with winding number $W_3=1$.
As discussed in the main text, we  consider the case with $m_0=4t_0$, $t_{\rm so}=0.5t_0$, and $\eta=0.2t_0$.
The dynamical field ${\bm g}({\bm k}) = (g_0, g_1, g_2, g_3)$ on the constant $g_0 = 0.5$ surface is shown in the upper panel of Fig.\ref{figS1}(c), constructed from $\widetilde{\langle\gamma_i ({\bm k})\rangle}$ ($i=1,2,3$) depicted in Fig.\ref{figS1}(d). The lower panel of Fig. \ref{figS1}(c) further illustrates ${\bm g}({\bm k})$ on the $g_0 = -0.5$ surface. These results demonstrate that on each constant-$g_0$ surface, the dynamical field exhibits a similar hedgehog configuration, indicating a nonzero winding number $W_3 = 1$.

\subsection{D.\quad Open boundary conditions and generalized Brillouin zone}

In the presence of skin effects, the energy spectrum of a NH system under open boundary conditions (OBCs) exhibits marked differences compared to the spectrum under periodic boundary conditions (PBCs); notably, the OBC spectrum can even be entirely real, as exemplified by the NH SSH model [Fig.~3(b)] discussed in the main text. The long-term dynamics resulting from complex and real spectra diverge significantly. For complex eigenenergies, only the right eigenvectors with positive imaginary parts (${\rm {Im}}(E)>0$) persist, while a real spectrum is characterized by the absence of gain or dissipation, resulting in dynamics akin to that of Hermitian systems. In the following, we will consider the 1D non-Bloch Hamiltonian of the form
\begin{align}
H(k_G) = h_x(k_G)\sigma_x+ h_y(k_G)\sigma_y,
\end{align}
and explore the quench dynamics associated with both the complex and real spectra in detail.
The non-Bloch Hamiltonian $H(k_{G})$ is obtained from $H(k)$ by the replacement $e^{\ui k}\to\beta\equiv e^{\ui k_{G}}$, where $k_{G}$ is complex~\cite{Yao2018_S}.
The corresponding $Q$ matrix $Q(k_G)$ also takes the form of Eq.~\eqref{Qk1}, i.e.,
\begin{align}
Q(k_G)=\frac{1}{2}\left[\frac{H(k_G)}{\varepsilon(k_G)}+\frac{H^\dag(k_G)}{\varepsilon^*(k_G)}\right],
\end{align}
with $E_\pm(k_G)=\pm \varepsilon(k_G)$ being the right eigenvalues of $H(k_G)$. Note that the proof presented in Eq.~\eqref{Hsproof} also applies to OBCs.

Here we show how to derive spin textures in the GBZ from a real-space initial state. 
The two complex-momentum bases are $\left\{\langle\beta^{\rm L}(k_G)|\right\}$ and $\left\{|\beta^{\rm R}(k_G)\rangle\right\}$,
where $|\beta^{\rm R}(k_G)\rangle=L^{-1/2}\sum_i e^{ik_G x_i}\ket{x_i}$ and $|\beta^{\rm L}(k_G)\rangle=L^{-1/2}\sum_i e^{ik_G^* x_i}\ket{x_i}$, with $i$ denoting the lattice site index and $L$ being the lattice length. We first consider the simple case where the complex-momentum bases are biorthogonal, i.e., they satisfy $\langle \beta^{\rm L}(k_G)|\beta^{\rm R}(k'_G)\rangle=\delta_{k_G,k'_G}$ and $\sum_{k_G}|\beta^{\rm R}(k_G)\rangle\langle \beta^{\rm L}(k_G)|=\id$.
The projection operator $P_{k_G}\equiv|\beta^{\rm R}(k_G)\rangle\langle \beta^{\rm L}(k_G)|$ can be introduced
to project the time-evolving density matrix $\rho(x,t)=e^{-\ui H t}\rho_0(x)e^{\ui H^\dag t}$, where $\rho_0(x)=\ket{\psi_0(x)}\bra{\psi_0(x)}$ with $\bra{\psi_0(x)}$ being the real-space initial state, 
onto the GBZ, leading to $\rho(k_G,t)=P_{k_G}\rho(x,t)P_{k_G}^\dag$.
This then gives the normalized LTSTs in the GBZ:
\begin{equation}~\label{LTST_GBZ_def}
\overline{\langle\sigma_i(k_G)\rangle}=\frac{{\langle\sigma_i(k_G)\rangle}_\infty}{\sqrt{\sum_{j=x,y} \langle\sigma_j(k_G)\rangle_\infty^2}} ,  \quad  i=x,y,
\end{equation}
where $\langle\sigma_i(k_G)\rangle_\infty=\lim_{t\to \infty}{\cal N}_{k_G}^{-1}{\rm Tr}[\rho(k_G,t)\sigma_i]$ with ${\cal N}_{k_G}\equiv{\rm Tr}[\rho(k_G,t)]$.

The projection procedure above can also be extended to more general scenarios where the complex-momentum bases are not biorthogonal.
Assuming that the time-evolving state $\ket{\psi(t)}=e^{-\ui H t}\ket{\psi_0}$ can be written as
$\ket{\psi(t)}=\sum_{k_G,\sigma=\uparrow\downarrow}C_{k_G}^\sigma(t)|\beta^{\rm R}(k_G),\sigma\rangle$,
we have
\begin{equation}
\langle\beta^{\rm L}(k_G),\sigma\ket{\psi(t)}=\sum_{k'_G} C_{k'_G}^\sigma(t) \langle\beta^{\rm L}(k_G)|\beta^{\rm R}(k'_G)\rangle, \quad \sigma=\uparrow\downarrow,
\end{equation}
or, in a matrix form,
\begin{equation}
{\bf D}^\sigma={\bf M}\cdot {\bf C}^\sigma,
\end{equation}
where the $N\times N$ matrix ${\bf M}$ is defined as ${\bf M}=\left[\langle\beta^{\rm L}(k_{G,i})|\beta^{\rm R}(k'_{G,j})\rangle\right]_{1\leq i,j\leq N}$, and the two vectors are ${\bf C}^\sigma=\left(C_{k_{G,1}}^\sigma,C_{k_{G,2}}^\sigma,\dots, C_{k_{G,N}}^\sigma\right)^{\rm T}$ and ${\bf D}^\sigma=\left(\langle\beta^{\rm L}(k_{G,1}),\sigma\ket{\psi(t)},\langle\beta^{\rm L}(k_{G,2}),\sigma\ket{\psi(t)},\dots, \langle\beta^{\rm L}(k_{G,N}),\sigma\ket{\psi(t)}\right)^{\rm T}$. Here $N$ is the number of lattice sites.
Therefore, for each spin $\sigma=\uparrow\downarrow$, the coefficients $C_{k_G}^\sigma(t)$ can be derived as follows:
\begin{equation}
{\bf C}^\sigma={\bf M}^{-1}{\bf D}^\sigma.
\end{equation}
The density matrix $\rho(t)$ is then given by $\rho(t)=\ket{\psi(t)}\bra{\psi(t)}=\sum_{k_Gk_G',\sigma\sigma'}C_{k_G}^{\sigma}(t)C_{k_G'}^{\sigma'}(t)|\beta^{\rm R}(k_G),\sigma\rangle\langle\beta^{\rm R}(k'_G),\sigma'|$. With this result, the normalized LTSTs $\overline{\langle\sigma_{x,y}(k_G)\rangle}$ can also be obtained as defined in Eq.~\eqref{LTST_GBZ_def}.

For a complex spectrum, following the procedure described in Eq.(\ref{LTST0}) and Eq.(\ref{LTST1}), one can prove that the normalized LTSTs satisfy
\begin{align}~\label{ResultsGBZ}
	\overline{\langle\sigma_i({k_G})\rangle} =
	\left\{ \begin{array}{ll}
		\overline{Q_i({k_G})} & \mathrm{Im}[\varepsilon({k_G})]>0\\
		-\overline{Q_i({k_G})} & \mathrm{Im}[\varepsilon({k_G})]<0
	\end{array} \right.,\quad i=x,y.
\end{align}
By defining the dynamical field ${\bm g}(k_G)=(g_x(k_G),g_y(k_G))$ as Eq.~\eqref{gi_defGen} with the replacement $k\rightarrow k_G$, one can then characterize the topology of the Hermitian matrix $Q(k_G)$ (and hence $H(k_G)$) by the winding number of ${\bm g}(k_G)$:
\begin{align}~\label{W1GBZ}
	W_1=\frac{1}{2\pi}\int_{\rm GBZ} d{k_G}\left[g_x(k_G)\partial_{{k_G}} g_y(k_G) - g_y(k_G)\partial_{{k_G}}g_x(k_G)\right].
\end{align}

For a real spectrum, we rewrite $H(k_G)$ as
\begin{align}
{h}_x(k_G) = {h}_{xr}(k_G)+\ui{h}_{xi}(k_G), \quad  {h}_y(k_G) = {h}_{yr}(k_G)+\ui{h}_{yi}(k_G),
\end{align}
where $h_{\alpha r}$ ($h_{\alpha i}$) represent the real (imaginary) parts of $h_{\alpha=x,y}$. Accordingly, the $Q$ matrix can be written as
\begin{equation}
	Q(k_G) =\frac{1}{\varepsilon(k_G)} ({h}_{xr}\sigma_{x} + {h}_{yr}\sigma_{y}).
\end{equation}
The spectrum being real requires $h_{yi}/h_{xr}= -h_{xi}/h_{yr}=q$ and $q^2<1$, which gives
$\varepsilon(k_G) = \sqrt{(h_{xr}^2+h_{yr}^2) (1-q^2)}$. 
Due to the bulk-surface duality~\cite{zhang2018dynamical_S},
the topological characterization of Hermitian Hamiltonians can reduce to topological invariants defined in particular momentum subspaces called band inversion surfaces (BISs).
Here, for $Q(k_G)$, the BIS refers to the complex momenta where $h_{xr}(k_G)=0$, and the associated topological invariant is given by
\begin{align}\label{C0}
	C_0=\frac{1}{2}\sum_{{\rm BIS}_i}\left[ \mathrm{sgn}[h_{yr}(k_{G,iR})]-\mathrm{sgn}[h_{yr}(k_{G,iL})]\right],
\end{align}
where $k_{G,iL/R}$ are a pair of $k_G$ that composes the $i$th BIS, i.e., $h_{xr}(k_{G,iL/R})=0$.

To characterize the topology of $Q(k_G)$, we introduce the time-averaged spin textures (TASTs)
\begin{align}
\langle\sigma_i(k_G)\rangle_T=\lim_{T\to \infty}\int_0^T\frac{1}{{\cal N}_{k_G}}{\rm Tr}\left[\rho(k_G,t)\sigma_i\right] dt,\quad i=x,y,
\end{align}
with the normalization factor ${\cal N}_{k_G}\equiv{\rm Tr}\left[\rho(k_G,t)\right]$.
When choosing the initial spin state to be $(\id -\sigma_x)/2$, one can derive that
\begin{equation}
	\begin{split}
		{\langle\sigma_x(k_G)\rangle}_T &= -1 + C\left(\frac{q^2h_{xr}^2+h_{yr}^2}{\varepsilon} \right), \\
		{\langle\sigma_y(k_G)\rangle}_T &= \lim_{T\to \infty}\int_0^T\frac{1}{{\cal N}_{k_G}} \left[\sin2\varepsilon t \frac{qh_{xr}}{ \varepsilon}-(1-\cos2\varepsilon t)\frac{h_{xr}{h}_{yr}}{h_{xr}^2+{h}_{y_r}^2}\right]dt,
	\end{split}
\end{equation}
where $C =\lim_{T\to \infty}\int_0^T\frac{1}{{\cal N}_{k_G}}(1-\cos 2\varepsilon t)dt$. Note that near ${h}_{xr}=0$,
\begin{equation}\label{GBZReal}
{\langle\sigma_y(k_G)\rangle}_T =-C \frac{{h}_{xr}{h}_{yr}}{{h}_{xr}^2+{h}_{y_r}^2},
\end{equation}
which indicate
\begin{equation}
\mathrm{sgn} [\partial_{k_{G}}{\langle\sigma_{y}(k_{G,j})\rangle}_T] = -\mathrm{sgn}[h_{yr}(k_{G,j})]\cdot\mathrm{sgn}[\partial_{k_{G}}h_{xr}(k_{G,j})],
\end{equation}
where $k_{G,j}$ can stand for $k_{G,iL}$ or $k_{G,iR}$. 
Since $\mathrm{sgn}[\partial_{k_{G}}h_{xr}(k_{G,jL})]=-\mathrm{sgn}[\partial_{k_{G}}h_{xr}(k_{G,jR})]$, we finally arrive at the conclusion
\begin{align}
C_0=\frac{1}{2}\sum_{j}\mathrm{sgn}\left[\partial_{k_G} {\langle\sigma_{y}(k_{G,j})\rangle}_T\right],
\end{align}
where $k_{G,j}$ denote the paired points at which ${\langle\sigma_{y}(k_{G,j})\rangle}_T=0$ and $|{\langle\sigma_{x}(k_{G,j})\rangle}_T|$ takes local minima.

Although our discussion in this subsection is restricted to 1D cases, the extension to higher-dimensional GBZ is straightforward. 
The key idea behind the dynamical characterization of 1D non-Bloch topological bands is to introduce the projection operator $P_{k_G}$, which is constructed using the complex-momentum bases $\{\langle\beta^{\rm L}(k_G)|\}$ and $\{|\beta^{\rm R}(k_G)\rangle\}$ introduced in the 1D GBZ theory~\cite{Yao2018_S}.
This operator projects the time-evolving wave function from real space into complex-momentum space [see the discussion above Eq.~\eqref{LTST_GBZ_def}]. 
Through this approach, we can extract the spin textures measured in the GBZ and use them to characterize the topology. Similarly, in higher odd dimensions
$d\geq3$, once the complex-momentum bases $\{\langle\beta_i^{\rm L}(k_G)|\}$ and $\{|\beta_i^{\rm R}(k_G)\rangle\}$ ($i=1,2\dots,d$) of the $d$-dimensional GBZ are determined through an appropriate theoretical framework, such as the recently-established Amoeba formulation~\cite{Wang2024_S} or geometry-adaptive spectral potential theory~\cite{Xiong2024_S},
one can dynamically measure the non-Bloch topological invariants by applying the same projection procedure.

\section{II.\quad Non-Hermitian systems with chiral symmetry}

\subsection{A.\quad Dynamical measurement scheme}

Chiral symmetry, which requires $\Gamma H^\dag({\bm k})\Gamma^{-1}= -H({\bm k})$, is notably distinct from sublattice symmetry in NH systems~\cite{Kawabata2019_S}. Here, we consider odd $d$D line-gapped NH systems, and write the Hamiltonian as
\begin{align}~\label{General_H_CS}
	\begin{split}
		H({\bm k})&= H_0({\bm k})+\ui H_1({\bm k}),\quad H_0({\bm k}) ={\sum}_{i=0}^d h_i({\bm k})\gamma_i, \quad H_1(\bm{k})= h'_{\Gamma}(\bm{k})\Gamma,
	\end{split}
\end{align}
with $\Gamma = \ui^{(d+1)/2} \prod_{i=0}^d \gamma_i$. We suppose that the non-Hermitian term ${h'}_{\Gamma}$ is sufficiently weak, allowing this system to exhibit real spectra $E_\pm(\bm{k})=\pm\varepsilon({\bm k})$ with $\varepsilon(\bm{k}) =\sqrt{\sum_i^d h_i^2-{{h}'_{\Gamma}}^2}$.
Consequently, the $Q$ matrix, defined in Eq.~\eqref{Q_Gen}, is now given by
\begin{align}
Q({\bm k})=\sum_iQ_i({\bm k})\gamma_i,\,\, {\rm where}\,\, Q_i(\bm{k})=\frac{h_i(\bm{k})}{\varepsilon(\bm{k})}.
\end{align}

To characterize the topology of $Q(\bm{k})$, we adopt the dynamical bulk-surface correspondence \cite{zhang2018dynamical_S}, which states that the bulk topology of a $d$D phase has a one-to-one correspondence to quench-induced dynamical topological patterns emerging on $(d-1)$D BISs. 
We introduce the TASTs
\begin{align}
	\langle\gamma_i(\bm{k})\rangle_T=\lim_{T\to \infty}\int_0^T\frac{1}{{\cal N}_{\bm{k}}}{\rm Tr}\left[e^{-\ui Ht}\rho_0 e^{\ui H^{\dagger}t} \gamma_i \right] dt,\,\,\, i=0,1,\dots,d
\end{align}
with the normalized factor ${\cal N}_{\bm{k}}\equiv{\rm Tr}\left[e^{-\ui Ht}\rho_0 e^{\ui H^{\dagger }t}\right]$.
The integrand can be rewritten as
\begin{equation}
	\begin{aligned}
		{\rm Tr}\left[e^{-\ui Ht}\rho_0 e^{\ui H^{\dagger}t} \gamma_i \right]={\rm Tr} \left[\rho_0\gamma_i \left(\cos^2 \varepsilon t - \ui \sin2\varepsilon t \frac{H_0}{\varepsilon}
		-\sin^2\varepsilon t \frac{\varepsilon^2 - 2\ui {h'}_{\Gamma} \Gamma H_0 + {h'}_{\Gamma}^2}{\varepsilon^2}+ \ui \sin2\varepsilon t \frac{h_i \gamma_i}{\varepsilon} + 2\sin^2\varepsilon t \frac{h_i \gamma_i H}{\varepsilon^2} \right)
		\right].
	\end{aligned}
\end{equation}

We first consider higher-dimensional systems with $d\geq3$ and set the initial state to be $\rho_0=(\id-\gamma_i)/2$ at each ${\bm k}$ ($i=0,1,\dots,d\,$). Without loss of generality, we choose $\rho_0=(\id-\gamma_0)/2$ in the following.
The TASTs are simplified to
\begin{equation}~\label{ATSTCS}
	{\langle\gamma_i(\bm{k})\rangle}_T =- C\frac{h_0({\bm k})h_i({\bm k})}{\varepsilon^2(\bm{k})},\quad i=1,2,\dots,d.
\end{equation}
Here $ C =\lim_{T\to \infty}\int_0^T\frac{1}{{\cal N}_{\bm{k}}} (1-\cos 2\varepsilon t)$ is positive.
We then define
\begin{align}~\label{BIS_def}
{\rm BIS}\equiv\{{\bm k}| h_0({\bm k})=0\}. 
\end{align} 
and introduce a dynamical spin-texture field $\boldsymbol{\mathcal{G}} (\bm{k})=(\mathcal{G}_1,\mathcal{G}_2,\dots,\mathcal{G}_d)$~\cite{zhang2018dynamical_S}, whose components are given by
\begin{equation}
	\label{eq.12}
	{\mathcal{G}_i(\bm{k})} \equiv -\frac{1}{\mathcal{N}_g}\partial_{k_\perp} {\langle\gamma_{i}\rangle}_T
\end{equation}
with $\mathcal{N}_g$ being the normalization factor and $k_\perp$ is the momentum perpendicular to BISs and pointing from the region ${h}_0<0$ to ${h}_0>0$. It can be derived from Eq.(\ref{ATSTCS}) that on the BIS,
\begin{equation}
	{\mathcal{G}_i(\bm{k})}|_{{\bm k}\in \mathrm{BIS}} = \overline{Q_i(\bm{k})},
\end{equation}
with $\overline{Q_i(\bm{k})}={Q_i(\bm{k})}/\sqrt{\sum_j Q_j(\bm{k})^2}$. Therefore, due to the topological equivalence between $H$ and $Q$, the line-gap topology of NH Hamiltonian can be characterized by a $(d-1)$D Chern number
\begin{equation}
	C_{d-1} = \sum_j \frac{\tilde{\Gamma}(d/2)}{2\pi^{d/2}}\frac{1}{(d-1)!}\int_{{\rm BIS}_j}{\boldsymbol{\mathcal{G}}}(\ud{\boldsymbol{\mathcal{G}}})^{d-1}
\end{equation}
where $\tilde{\Gamma}(x)$ is the Gamma function, ${\boldsymbol{\mathcal{G}}}(\ud{\boldsymbol{\mathcal{G}}})^{d}\equiv\epsilon^{i_{1}i_{2}\cdots i_{d+1}}\mathcal{G}_{i_{1}}\ud \mathcal{G}_{i_{2}}\wedge\cdots\wedge\ud \mathcal{G}_{i_{d+1}}$
with $\epsilon^{i_{1}i_{2}\cdots i_{d+1}}$ being the fully antisymmetric tensor and $i_{1,2,\dots,d+1}\in\{0,1,2,\dots,d\}$,
and `$\ud$' denotes the exterior derivative.

However, for 1D systems, the dynamical characterization diverges significantly.
Rather than being polarized in $\gamma_0$, the initial state should be chosen as $\rho_0 = (\id-\Gamma)/{2}$. Suppose that $\gamma_0=\sigma_x$, $\gamma_1=\sigma_y$ and $\Gamma=\ui\sigma_x\sigma_y=-\sigma_z$. We have $\rho_0 = (\id+\sigma_z)/{2}$ and
\begin{align}~\label{TAST_1D}
{\langle\sigma_x\rangle}_T= -C_1 \frac{h_y}{\varepsilon^2}, \quad {\langle\sigma_y\rangle}_T = C_1 \frac{h_x}{\varepsilon^2},
\end{align}
where $C_1 =\lim_{T\to \infty}\frac{1}{T}\int_{0}^{T}dt  {\cal N}_{k}^{-1} \left[-h'_z+h'_z \cos(2\varepsilon t)-\varepsilon\sin(2\varepsilon t)\right]$.
We employ the normalized TASTs
\begin{equation}~\label{Ave_def}
	\overline{\langle\sigma_i(k)\rangle}_T=\frac{{\langle\sigma_i(k)\rangle}_T}{\sqrt{\sum_{j=x,y} \langle\sigma_{j}(k)\rangle_T^2}} ,\quad i =x,y,
\end{equation}
which define the dynamical field ${\bm g}(k)=(g_{x}(k),g_{y}(k))$ as
\begin{equation}
	\begin{split}
		g_{x} = \overline{\langle\sigma_y({k})\rangle}_T,\quad
		g_{y} =-\overline{\langle\sigma_x({k})\rangle}_T.
	\end{split}
\end{equation}
The topology of the Hermitian $Q$ matrix is characterized by the winding number of the dynamical field ${\bm g}(k)$:
\begin{align}~\label{W1CS}
	W_1=\frac{1}{2\pi}\int_{\rm BZ} d{k}\left[g_{x}(k)\partial_{{k}} g_{y}(k)-g_{y}(k)\partial_{{k}}g_{x}(k)\right].
\end{align}

\begin{figure*}
	\includegraphics[width=0.99\textwidth]{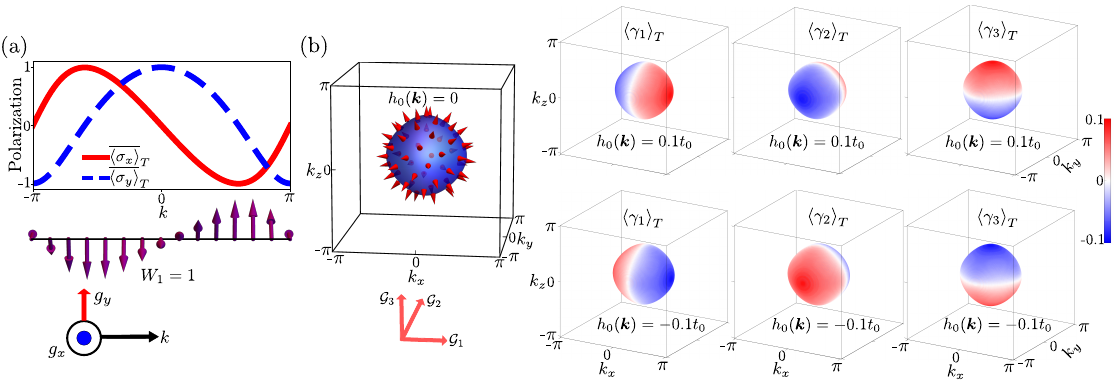}
	\caption{Measuring the topology of NH Hamiltonian with chiral symmetry. 
		(a) 1D NH SSH model. The dynamical field $\bm{g}(k)$ (purple arrows), constructed from $\overline{\langle \sigma_{x,y} \rangle}_T$, exhibits the integer-valued winding $W_1 = 1$. Here $t_1=0.3t_2,\eta=0.5t_2$.  
		(b)  3D NH topological insulator. 
Left panel: The dynamical spin-texture field $\boldsymbol{\mathcal{G}} (\bm{k})=(\mathcal{G}_1,\mathcal{G}_2,\mathcal{G}_3)$ (arrows) exhibits a hedgehog configuration on the BIS where $h_0(\bm{k}) = 0$, yielding $C_2=1$. 
Right panel: The field components $\mathcal{G}_i(\bm{k})$ ($i=1,2,3$) are constructed by subtracting the values of ${\langle \gamma_{i}(\bm{k})\rangle}_T$ on the $h_0=\pm0.1t_0$ surfaces.
Here, $m_0=4t_0$, $t_{\rm so}=0.5t_0$ and $\eta=0.2t_0$.
	}\label{figS2}
\end{figure*}

\subsection{B.\quad 1D and 3D examples}

We first consider the 1D NH SSH model with spin-dependent dissipation, described by the Hamiltonian
\begin{equation}
	\begin{split}
		H(k) =(t_1+t_2\cos k)\sigma_x+t_2\sin k\sigma_y -\ui\eta\sigma_z.
	\end{split} 
\end{equation}
We assume $\eta<\sqrt{t_1^2+t_2^2+ 2t_1t_2\cos k}$ such that the system is line-gapped and exhibits a real spectrum. 
This NH system is topologically nontrivial (trivial) when $t_1<t_2$ ($t_1>t_2$).
The NH topology can be dynamically detected using the procedure detailed in Eqs.~\eqref{TAST_1D}-\eqref{W1CS}.
An example with $t_1=0.3t_2$ and $\eta=0.5t_2$ is shown in Fig.~\ref{figS2}(a).
The nonzero winding of the dynamical field ${\bm g}(k)$, constructed from the normalized TASTs $\overline{\langle\sigma_{x,y}(k)\rangle}_T$, characterizes the line-gap topology, yielding $W_1=1$.

We also present an example of 3D NH topological insulators with CS. The Hamiltonian takes the form of Eq.~\eqref{General_H_CS}, with the Hermitian component $H_0$ corresponding to that of the 3D model discussed in the main text, i.e., $h_0= m_0 -2t_0\sum_{i=1}^{3} \cos k_{r_i}$, $h_{i>0}=2t_{\rm so}\sin k_{r_i}$, and $h_\Gamma'=\eta$.
The system features a real spectrum when $\eta<\sqrt{\sum_{i=0}^3 h_i^2}$, and is topologically nontrivial in the range $-6t_0<m_0<6t_0$ (cf. Sec.~\ref{SecIB}). 
Taking the parameters $m_0=4t_0$, $t_{\rm so}=0.5t_0$, and $\eta=0.2t_0$ as an example, we present the calculated spin textures ${\langle \gamma_{i} \rangle}_T$ ($i=1,2.3$) on two constant-$h_0$ surfaces, positioned slightly inside and outside the BIS ($h_0(\bm{k})=0$), respectively, in the right panel of Fig.~\ref{figS2}(b). The direction of the field $\boldsymbol{\mathcal{G}}(\bm{k})$ on the BIS is determined by subtracting ${\langle \gamma_{i} \rangle}_T$ values from adjacent constant-$h_0$ surfaces. By analyzing the winding of  $\boldsymbol{\mathcal{G}}(\bm{k})$, one can derive the bulk topological invariant: $W_3=C_2=1$, as illustrated in the left panel of Fig.~\ref{figS2}(b).

\section{III.\quad Summary of dynamical characterization methods}

To summarize, we provide an overview of dynamical characterization methods employed in our measurement scheme for systems under PBCs. As emphasized throughout the discussion, a carefully selected initial state can significantly simplify the measurement process. 
These choices are closely tied to the dimensionality and symmetry of the non-Hermitian system under consideration.
A summary of the selected initial states is provided in Table~\ref{TableSI}.

The differentiation in initial-state choices and methods stems from the fact that both the number and dimensionality of the $\gamma$ matrices that constitute our model Hamiltonian 
[refer to Eqs.~\eqref{General_H_SLS} and \eqref{General_H_CS}] strongly depend on the system’s dimensionality. 
In particular, 1D systems are characterized by only three Pauli matrices and two energy bands. 
The minimal degrees of freedom governing the quench dynamics necessitate the use of distinct initial states or even different characterization 
methods for 1D systems as compared to higher-dimensional systems.

Furthermore, unlike systems with SLS, chiral-symmetric systems exhibit real energy spectra under conditions of weak dissipation. 
This results in quench dynamics that remain unitary, rendering the situation analogous to the characterization of Hermitian systems~\cite{zhang2018dynamical_S}. 
In such cases, spin oscillations on BISs are resonant and pronounced, while spin dynamics away from the BISs become negligible.
As a result, in higher-dimensional systems with chiral symmetry, dynamical characterizations are primarily based on measurements taken on the BISs 
rather than across the entire BZ.
Table~\ref{TableSI} also provides a summary of how to construct the dynamical field.

\begin{table}[ht]
	\centering
	\begin{tabular}{|c|c|c|c|c|c|}
		\hline
		Dimensionality & Symmetry & Energy Spectrum & Initial State & \multicolumn{2}{|c|}{\,Dynamical Field\,}\\
		\hline
		\multirow{2}{*}{$d=1$} & SLS & Complex & Arbitrary &\,\, LTST \,\,& BZ \\
		\cline{2-6}
		&  CS & Real & $\rho_0=(\id-\Gamma)/2 $ & TAST & BZ \\
		\hline
		\multirow{2}{*}{$d\geq3$} & SLS & Complex & $\rho_0=(\id-S)/2 $ & LTST & BZ\\
		\cline{2-6}
		& CS & Real & $\,\rho_0=(\id - \gamma_i)/2$,\, $i=0,1,\dots,d\,$ & TAST & BIS\\
		\hline
	\end{tabular}
	\caption{Summary of dynamical characterization methods for odd-dimensional systems under PBCs, categorized by dimensionality and symmetry.
	Chiral-symmetric systems are considered under conditions of weak dissipation.
	Here, $\Gamma$ and $S$ represent the operators of chiral symmetry (CS) and sublattice symmetry (SLS), respectively. 
	Initial states are all assumed to be uniformly distributed in quasimomentum space, and thus explicitly represented as spin states for simplicity.
	The column labeled `Dynamical Field' specifies whether the dynamical field ${\bm g}(\bm{k})$ is constructed from long-time spin textures (LTSTs) or time-averaged spin textures (TASTs) and whether it is defined across the entire Brillouin zone (BZ) or restricted to band-inversion surfaces (BISs).
	}
	\label{TableSI}
\end{table}

\section{IV.\quad Experimental realization and detection in ultracold atoms}

\subsection{A.\quad Experimental realization with $^{40}$K atoms}

\begin{figure}
	\includegraphics[width=0.5\textwidth]{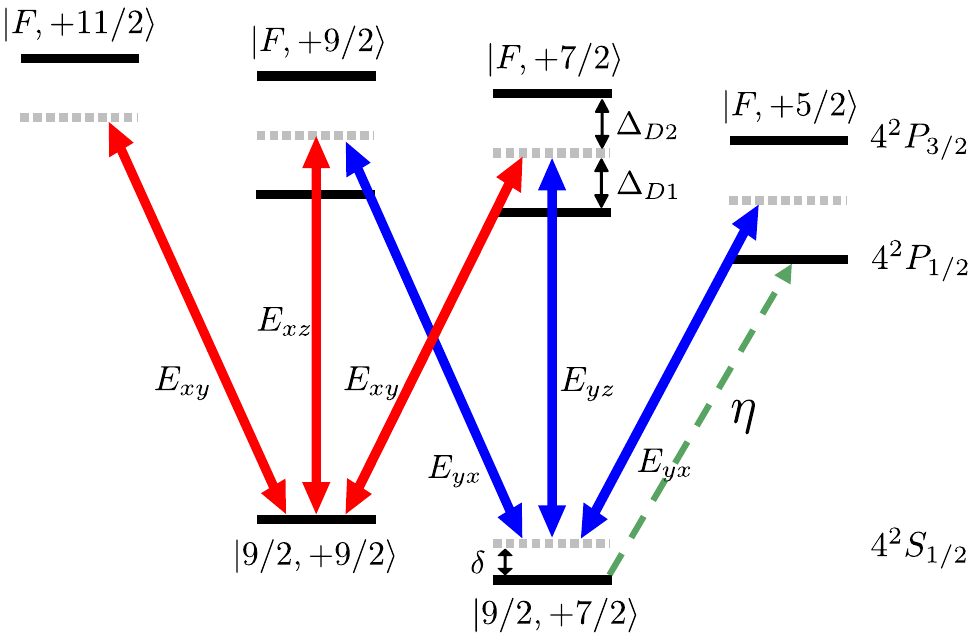}
	\caption{Laser couplings for $^{40}\rm{K}$ atoms. The wavelengths of the Raman and lattice beams are set to be between the $D_1$ and $D_2$ lines such that $\Delta_{D_1}>0$ and $\Delta_{D_2}<0$. A near-resonant beam (green) is added to induce a spin-dependent atom loss $\eta$.
	}\label{figS3}
\end{figure}

We take $^{40}$K atoms as an example to illustrate our experimental scheme while all the results are applicable to other atoms. 
For $^{40}$K, the spin-$1/2$ system can be constructed
by $\ket{\uparrow}=\ket{F=9/2, m_F=9/2}$ and $\ket{\downarrow}=\ket{9/2, 7/2}$.  As shown in Fig.~\ref{figS3}, the lattice and Raman coupling potentials are contributed from both the $D_2$ ($4{^{2}S}_{1/2}\to4{^{2}P}_{3/2}$) and $D_1$ ($4^{2}S_{1/2}\to4^{2}P_{1/2}$) lines.
Here, we only consider the Hermitian part of the total Hamiltonian, which reads ($\hbar=1$)
\begin{align}\label{Ham_S}
	H_0=\left[\frac{k_x^2}{2m}+V(x)\right]\otimes\id+{\cal M}_x(x)\sigma_x+{\cal M}_y(x)\sigma_y+m_z\sigma_z,
\end{align}
where $V(x)$ denotes the lattice potential, ${\cal M}_{x,y}(x)$ are Raman potentials, and $m_z=\delta/2$ measures the two-photon detuning.

As illustrated in Fig.~4 in the main text, the light fields take the form
\begin{align}
	\begin{split}
		{\bm E}_x&=\hat{y}E_{xy}e^{\ui(\alpha+\alpha_L/2)}\cos(k_0x-\alpha_L/2)+\ui\hat{z}E_{xz}e^{\ui(\alpha+\alpha_L/2)}\sin(k_0 x-\alpha_L/2),\\
		{\bm E}_y&=\hat{x}E_{yx}e^{\ui(k_0y+\beta)}+\ui\hat{z}E_{yz}e^{\ui(k_0y+\beta)}.
	\end{split}
\end{align}
where $\alpha$ and $\beta$ denote the initial phases, and $\alpha_L$ is the phase acquired by ${\bm E}_x$ for an additional optical path back to the atom cloud.
Two independent Raman transitions are driven by the components $E_{xz}, E_{yx}$ and $E_{xy},E_{yz}$ , respectively, which leads to
\begin{align}
	M_1=\sum_F\frac{\Omega^{(3/2)*}_{\uparrow F,x-}\cdot\Omega^{(3/2)}_{\downarrow F,yz}}{\Delta_{D_2}}+\sum_F\frac{\Omega^{(1/2)*}_{\uparrow F,x-}\cdot\Omega^{(1/2)}_{\downarrow F,yz}}{\Delta_{D_1}},\quad
	M_2=\sum_F\frac{\Omega^{(3/2)*}_{\uparrow F,xz}\cdot\Omega^{(3/2)}_{\downarrow F,y+}}{\Delta_{D_2}}+\sum_F\frac{\Omega^{(1/2)*}_{\uparrow F,xz}\cdot\Omega^{(1/2)}_{\downarrow F,y+}}{\Delta_{D_1}},
\end{align}
where $\Delta_{D_1}>0$, $\Delta_{D_2}<0$, $\Omega^{(J)}_{\sigma F,\mu z}=\langle\sigma|ez|F,m_{F\sigma},J\rangle\hat{e}_z\cdot{\bm E}_{\mu}$ and
$\Omega^{(J)}_{\sigma F,\mu\pm}=\langle\sigma|ex|F,m_{F\sigma}\pm1,J\rangle\hat{e}_\pm\cdot{\bm E}_{\mu}$ ($\mu=x,y$). Since $\hat{e}_\pm=\mp(\hat{x}\pm\ui\hat{y})/\sqrt{2}$, we have $\hat{e}_+\cdot{\bm E}_{y}=-E_{yx}/\sqrt{2}$ and $\hat{e}_-\cdot{\bm E}_{x}=\ui E_{xy}/\sqrt{2}$. From the dipole matrix elements of $^{40}$K~\cite{Tiecke2011_S}, we obtain
\begin{align}
	{\cal M}_x(x)=M_{0x}\cos(k_0 x-\alpha_L/2),\quad
	{\cal M}_y(y)=M_{0y}\sin(k_0 x-\alpha_L/2),
\end{align}
where
\begin{equation}\label{M0xy_S}
	M_{0x/0y}=\frac{t_{D_1}^2}{9}\left(\frac{1}{|\Delta_{D_1}|}+\frac{1}{|\Delta_{D_2}|}\right)E_{xy/xz}E_{yz/yx},
\end{equation}
with the transition matrix elements $t_{D_1}\equiv\langle J=1/2||e{\bf r}||J'=1/2\rangle$, $t_{D_2}\equiv\langle J=1/2||e{\bf r}||J'=3/2\rangle$ and $t_{D_2}\approx\sqrt{2}t_{D_1}$.
The optical lattice is given by
\begin{align}
	V_{\sigma=\uparrow,\downarrow}=\sum_{F,\zeta=z,\pm}\frac{\left|\Omega^{(3/2)}_{\sigma F,x\zeta}\right|^2}{\Delta_{D_2}}+\sum_{F,\zeta=z,\pm}\frac{\left|\Omega^{(1/2)}_{\sigma F,x\zeta}\right|^2}{\Delta_{D_1}}=V_{0}\cos^2(k_0 x-\alpha_L/2),
\end{align}
where
\begin{equation}
	V_{0}=\frac{t_{D_1}^2}{3}\left(\frac{1}{|\Delta_{D_1}|}-\frac{2}{|\Delta_{D_2}|}\right)(E_{xy}^2-E_{xz}^2).
\end{equation}

In the tight-binding limit and only considering $s$-bands, the Hamiltonian (\ref{Ham_S}) takes the form~\cite{LiuXJ2013_S,Wang2018_S} 
\begin{align}
\begin{split}
	H_0=&-t_0\sum_{\langle i_xj_x\rangle}\left(c^\dagger_{i\uparrow}c_{j\uparrow}-c^\dagger_{i\downarrow}c_{j\downarrow}\right)
	+\left[\sum_{i}t_{\rm so}(c^\dagger_{i\uparrow}c_{i+1\downarrow}-c^\dagger_{i\uparrow}c_{i-1\downarrow})+{\rm h. c.}\right] \\
	&-\ui m_y\sum_{i}(c^\dagger_{i\uparrow}c_{i\downarrow}-c^\dagger_{i\downarrow}c_{i\uparrow}) +\sum_{\vec{i}}m_z(n_{i\uparrow}-n_{i\downarrow}).
\end{split}
\end{align}
where $t_0=-\int dx\phi_{s}(x)\left[\frac{{\bf k}^2}{2m}+V_{\rm latt}(x)\right]\phi_{s}(x-a)$, $t_{\rm so}=M_{0x}\int dx\phi_{s}(x)\cos(k_0x)\phi_{s}(x-a)$, and $m_y=M_{0y}\int dx \phi_{s}(x)\sin(k_0x)\phi_{s}(x)$.
Here $\phi_{s}(x)$ denotes the Wannier function.
Through Fourier transformation, we have $H_0=\sum_{q} {\Psi}^\dagger_{q}{\cal H}_0(q){\Psi}_{q}$, where ${\Psi}_{q}=({c}_{q\uparrow},{c}_{q\downarrow})$ and
the Bloch Hamiltonian
\begin{align}\label{Hq}
{\cal H}_0(q)=(m_z-2t_0\cos qa){\sigma}_z+(m_y+2t_{\rm so}\sin qa){\sigma}_y.
\end{align}
Here $a$ is the lattice constant and $q$ is the quasimomentum.

\begin{figure*}
	\includegraphics[width=0.97\textwidth]{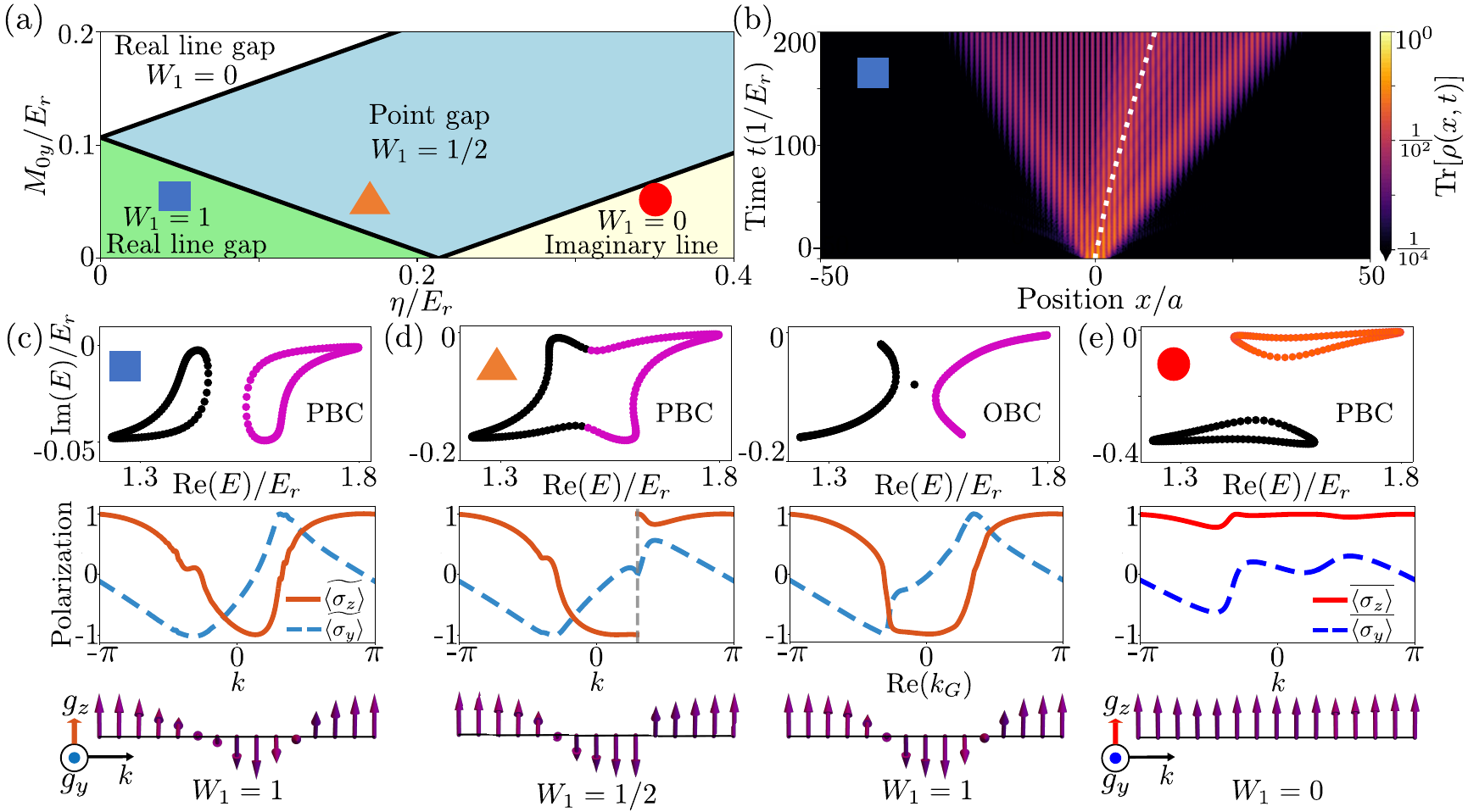}
	\caption{Numerical simulation of the experimental realization in cold atoms.
		(a) Phase diagram as function of $\eta$ and $M_{0y}$, obtained in the tight-binding limit under PBCs. 
		Here we take $(V_0,M_{0x},m_z)=(4,1,0.1)E_r$.
		(b) The wave-packet evolution in the real-line-gapped phase. The density distribution is displayed as a function of position $x$ and time $t$. 
		The white dotted line indicates the trajectory of the center of mass.
		(c)-(e) Dynamical measurement. Top: Complex-energy spectra. Middle: The spin textures measured after a duration of $\tau^*=200/E_r$ in (c), $\tau^*=100/E_r$ in (d) and $\tau^*=50/E_r$ in (e). Bottom: The constructed dynamical field ${\bm g}(k)$. Here we take $(\eta,M_{0y}) =(0.05,0.05)E_r$ in (c), $(0.17,0.05)E_r$ in (d) and $(0.35,0.05)E_r$ in (e)
	}\label{figS4}
\end{figure*}

\subsection{B.\quad Numerical results of dynamical measurements}

To demonstrate the experiment's feasibility, we simulate the quench dynamics based on the Hamiltonian (11) in the main text. Taking the experimentally viable parameters $(V_0,M_{0x},m_z)=(4,1,0.1)E_r$, we present the system's phase diagram as function of $\eta$ and $M_{0y}$ in Fig.~\ref{figS4}(a), derived in the tight-binding limit under PBCs. By setting $M_{0y}=0.05E_r$, we examine three distinct cases with $\eta=0.05E_r$, $0.17E_r$, and $0.35E_r$, represented by square, triangle, and circle markers in Fig.~\ref{figS4}(b). These cases correspond to real-line-gapped topological, point-gapped, and imaginary-line-gapped trivial phases, respectively,  all of which are measurable through quench dynamics. 

We prepare the initial states as Wannier states localized at the center of the lattice, shaped by a Gaussian envelope and polarized in the spin-up state. We investigate the time evolution under the Hamiltonian presented in Eq.~(11) of the main text. The results for the case in the real-line-gapped topological phase are displayed in Fig.~\ref{figS4} (b), illustrating how the density distribution evolves in space and time. The center of mass movement is indicated by a white dotted line, highlighting the presence of the skin effect.

After a finite time $\tau^*\gg1/\eta$, which is is typically much shorter than the cold-atom system's lifetime $\tau \sim 1300/E_r$ (about 150ms) for $^{40}$K atoms~\cite{Wang2018_S}, we project the final states onto the BZ or GBZ to obtain the normalized LTSTs. The results for the three cases are presented in Fig.~\ref{figS4}(c)-(e), where we account for the elimination of high-energy band effects and select the lattice points $x=na$ ($n={-N/2,\dots,N/2}$ and $N$ is the total number of lattice sites) during the projection. 
As shown in Fig.~\ref{figS4}(a), in the real-line-gapped case, the dynamical field ${\bm g}(k)$, constructed from the normalized LTSTs, indicates the nontrivial topology with $W_1=1$.
In the point-gapped case, we present the results under both PBCs and OBCs in Fig.~\ref{figS4}(b). The winding number $W_1=1/2$ with an abrupt $\pi$ jump signifies the unknot braiding of complex-energy bands under PBCs, while $W_1=1$ characterizes the existence of topological zero modes under OBCs. 
The result for the imaginary-line-gapped phase is shown in Fig.~\ref{figS4}(c), where the dynamical field is constructed from the TASTs and $W_1=0$ confirms that it is a trivial phase. 

Please note that, in addition to ensuring sufficient time for the dynamical measurement, it is crucial for the atom cloud to remain well away from the lattice boundary, as the boundary can reflect the wave-packet's motion and interfere with the measurements. 
Under this circumstance, the distinction between PBCs and OBCs in the measured results depends solely on how the measurements are projected. Specifically,
PBC (OBC) results are obtained by projecting the measurements onto the BZ (GBZ).
This requirement adds another constraint to the experiment. However, this is not an issue in our settings, as demonstrated in Fig.~\ref{figS4}(b), where the wave packet remains significantly distant from the boundary even after $\tau^* = 200/E_r$.

\subsection{C.\quad Mapping from the BZ to the GBZ}

In cold-atom experiments, measuring observables in quasi-momentum space, such as spin textures and energy dispersion $\varepsilon(k)$, is relatively straightforward. 
Here we demonstrate that by utilizing these observables measured in the BZ, one can construct the dynamical field ${\bm g}(k_G)$ defined in the GBZ
to characterize non-Bloch topology, thus circumventing the challenging of measuring the real-space wave function. 
In the following, we focus on 1D real-line-gapped systems with SLS, while the analysis can be extended to point-gapped or imaginary-line-gapped systems in a similar manner.

We assume the initial state $\ket{\psi_0(x)}$ is prepared at the lattice center with spin $\ket{\uparrow}$, and during the evolution, the wave function keeps far away from the boundary.
Projecting the initial state into quasi-momentum space gives $\rho_0= \sum_{kk'} p_{kk'} \ket{k,\uparrow}\bra{k',\uparrow}$ with $p_{kk'} = \bra{k}\ket{\psi_0(x)}\bra{\psi_0(x)}\ket{k'}$.
After time $t$, we have
\begin{align}~\label{rhot_BZ}
\rho(t) &= \sum_{kk'} p_{kk'} e^{-\ui H(k)t}|k,\uparrow\rangle\langle k',\uparrow\!|e^{\ui H^\dag(k)t}\nonumber\\
& = \sum_{kk'} p_{kk'}\left[\mathcal{A}_{kk'}(t)|k,\uparrow\rangle\langle k',\uparrow\!|+\mathcal{B}_{kk'}(t)\theta_k |k,\downarrow\rangle\langle k',\uparrow\!|+\mathcal{C}_{kk'}(t) \theta^*_{k'} |k,\uparrow\rangle\langle k',\downarrow\!| +\mathcal{D}_{kk'}(t)\theta_k\theta^*_{k'} |k,\downarrow\rangle\langle k',\downarrow\!| \right],
\end{align}
where $\theta_k = b_k/a_k$ represents the ratio of the spin components of the right eigenvector $|u^{\rm R}(k)\rangle = a_k \ket{k,\uparrow} + b_k \ket{k,\downarrow}$, and
\begin{equation}~\label{ABCD_def}
	\begin{aligned}
		&\mathcal{A}_{kk'}(t)= \frac{1}{4}\left(e^{-\ui[\varepsilon(k)-\varepsilon^*(k')]t} +  e^{\ui[\varepsilon(k)+\varepsilon^*(k')]t} +  e^{-\ui[\varepsilon(k)+\varepsilon^*(k')]t} +  e^{\ui[\varepsilon(k)-\varepsilon^*(k')]t}\right),\\
		&\mathcal{B}_{kk'}(t)= \frac{1}{4}\left(e^{-\ui[\varepsilon(k)-\varepsilon^*(k')]t} -e^{\ui[\varepsilon(k)+\varepsilon^*(k')]t} +  e^{-\ui[\varepsilon(k)+\varepsilon^*(k')]t} -e^{\ui[\varepsilon(k)-\varepsilon^*(k')]t}\right),\\
		&\mathcal{C}_{kk'}(t)= \frac{1}{4}\left(e^{-\ui[\varepsilon(k)+\varepsilon^*(k')]t} +e^{\ui[\varepsilon(k)+\varepsilon^*(k')]t} -  e^{-\ui[\varepsilon(k)+\varepsilon^*(k')]t} -e^{\ui[\varepsilon(k)-\varepsilon^*(k')]t}\right),\\
		&\mathcal{D}_{kk'}(t)= \frac{1}{4}\left(e^{-\ui[\varepsilon(k)+\varepsilon^*(k')]t} -e^{\ui[\varepsilon(k)+\varepsilon^*(k')]t} -  e^{-\ui[\varepsilon(k)+\varepsilon^*(k')]t} +e^{\ui[\varepsilon(k)-\varepsilon^*(k')]t}\right).
	\end{aligned}
\end{equation}

We first consider the simple case where the two complex-momentum bases are biorthogonal, i.e., $\langle \beta^{\rm L}(k_G)|\beta^{\rm R}(k'_G)\rangle=\delta_{k_G,k'_G}$. Here $|\beta^{\rm R}(k_G)\rangle=L^{-1/2}\sum_i e^{\ui k_G x_i}\ket{x_i}$ and $|\beta^{\rm L}(k_G)\rangle=L^{-1/2}\sum_i e^{\ui k_G^* x_i}\ket{x_i}$. The time-evolving wavefunction projected onto the complex-momentum bases can be written as 
$\rho(t) = \sum_{k_Gk'_G,\sigma\sigma'} p_{k_G k'_G}^{\sigma\sigma'}(t)|\beta^{\rm R}(k_G),\sigma\rangle \langle \beta^{\rm R}(k'_G),\sigma'|$.
 We define
 \begin{equation}~\label{Tkk_def}
T_{k_Gk} \equiv\langle\beta^{\rm L}(k_G) |k \rangle = \frac{1}{L} \sum_i e^{-\ui(k_G-k)x_i},
 \end{equation}
which gives
 \begin{equation}
 p_{k_G k_G}^{\sigma\sigma'}(t)=\sum_{k,k'}T_{k_Gk}p^{\sigma\sigma'}_{kk'}(t)T^*_{k_Gk'},
  \end{equation}
where
\begin{equation}~\label{Psskk}
  p^{\uparrow\uparrow}_{kk'}(t)= p_{kk'}{\cal A}_{kk'}(t), \quad p^{\downarrow\uparrow}_{kk'}(t)= p_{kk'}{\cal B}_{kk'}(t)\theta_k,\quad
 p^{\uparrow\downarrow}_{kk'}(t)= p_{kk'}{\cal C}_{kk'}(t)\theta^*_{k'}, \quad  p^{\downarrow\downarrow}_{kk'}(t)= p_{kk'}{\cal D}_{kk'}(t)\theta_k\theta^*_{k'}.
\end{equation}
One can obtain LTSTs in the GBZ as
\begin{equation}~\label{LTST_measure}
		 \langle \sigma_{x}(k_G)\rangle_\infty=\lim_{t\to\infty}\frac{p_{k_Gk_G}^{\uparrow\downarrow}(t)+ p_{k_Gk_G}^{\downarrow\uparrow}(t)}{p_{k_Gk_G}^{\uparrow\uparrow}(t)+ p_{k_Gk_G}^{\downarrow\downarrow}(t)},\quad
		\langle \sigma_{y}(k_G) \rangle_\infty = \lim_{t\to\infty} \ui \frac{p_{k_Gk_G}^{\uparrow\downarrow}(t) - p_{k_Gk_G}^{\downarrow\uparrow}(t)}{p_{k_Gk_G}^{\uparrow\uparrow}(t)+ p_{k_Gk_G}^{\downarrow\downarrow}(t)}, \quad
		 \langle \sigma_{z}(k_G)\rangle_\infty=  \lim_{t\to\infty}\frac{p_{k_Gk_G}^{\uparrow\uparrow}(t) - p_{k_Gk_G}^{\downarrow\downarrow}(t)}{p_{k_Gk_G}^{\uparrow\uparrow}(t)+ p_{k_Gk_G}^{\downarrow\downarrow}(t)}.
\end{equation}
As discussed in the above subsection, the limit $t\to\infty$ can be approximated as $t>\tau^*$, where $\tau^*$ denotes a finite time satisfying $\tau^*\gg 1/\eta$.
With the result in Eq.~\eqref{LTST_measure}, 
one can further derive both the normalized and deformed LTSTs as defined in Eq.~\eqref{LTST_def} and Eq.~\eqref{gi_defGen}, respectively.
The dynamical field ${\bm g}(k_G)$ can then be constructed to characterize the topology.

\begin{figure}
	\includegraphics[width=0.45\textwidth]{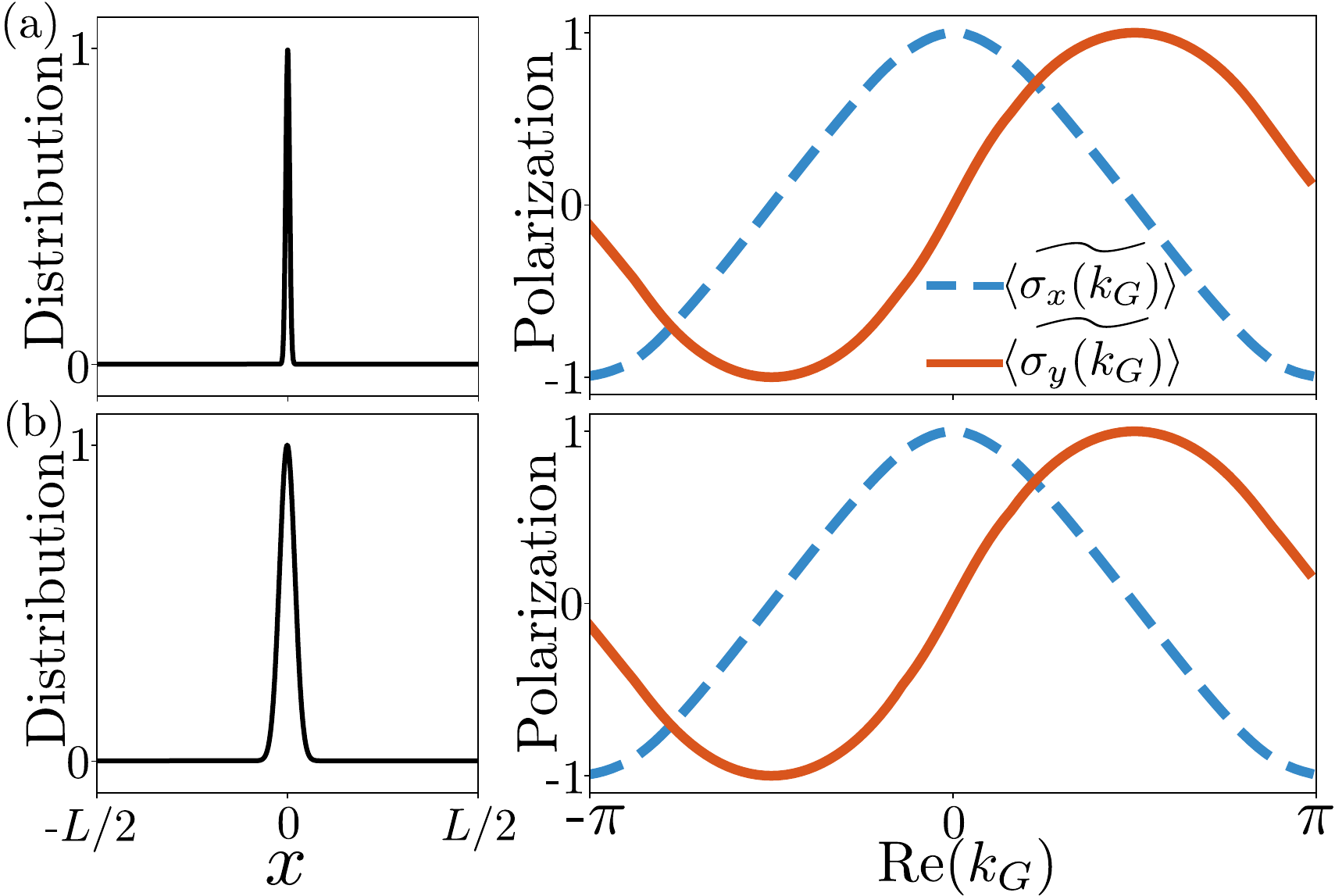}
	\caption{The construction of deformed LTSTs in the GBZ  with different initial states for the 1D NH SSH model. The parameters are $(t_1,\eta) = (0.3t_2,0.5t_2)$, consistent with those in Fig.~1(c) of the main text. Spin textures in the BZ are measured at time $t=10/t_2$. The initial state is modeled as a Gaussian wave packet  with a width of $d= L/200$ in (a) and $d=L/50$ in (b) with $L=100a$, where $a$ is the lattice constant and $L$ represents the lattice length.
	}\label{figS5}
\end{figure}

From the derivation above, it is clear that the crucial step lies in determining $p^{\sigma\sigma'}_{kk'}(t)$, as defined in Eq.~\eqref{Psskk}.
It is important to note that the quantities ${\cal A}_{kk'}$, ${\cal B}_{kk'}$, ${\cal C}_{kk'}$, ${\cal D}_{kk'}$ can be extracted directly from the measured energy spectrum using Eq.~\eqref{ABCD_def},
Additionally, the ratio $\theta_k$, which characterizes the spin components of the right eigenvector, can be directly measured in real-line-gapped systems as follows:
\begin{align}~\label{thetak_measure}
	\theta_k =
	\left\{ \begin{array}{ll}
		\frac{\langle \sigma_{x}(k)\rangle_\infty}{1+\langle \sigma_z(k) \rangle_\infty} + \ui\frac{\langle \sigma_{y}(k) \rangle_\infty}{1+\langle \sigma_z(k) \rangle_\infty}  & \mathrm{Im}[\varepsilon({k})]>0\\
		-\frac{\langle \sigma_{x}(k)\rangle_\infty}{1+\langle \sigma_z(k) \rangle_\infty} -\ui\frac{\langle \sigma_{y}(k) \rangle_\infty}{1+\langle \sigma_z(k) \rangle_\infty}  & \mathrm{Im}[\varepsilon({k})]<0
	\end{array} \right. ,
\end{align}
where $\langle \sigma_{x,y,z}(k)\rangle_\infty$ represent the spin textures measured in quasimomentum space at a sufficiently long time $t>\tau^*$.
The only quantity that cannot be determined through measurement is  $p_{kk'}$, which reflects the correlation between $k$ and $k'$ in the initial state.

However, we will demonstrate that, due to the nonunitary nature of the time evolution, the specific values of $p_{kk'}$ are not critical for characterizing the topology. 
To illustrate this, we consider the NH SSH model discussed in the main text. The initial state is assumed to take the form $\rho_0(x)=\mathcal{N}^{-1} e^{-x^2/2d^2}\ket{\uparrow}\bra{\uparrow}$, where $d$ denotes the width of the wave packet and $\mathcal{N}$ is the normalization factor.
It is important to note that the actual initial state used in our calculations for Fig. 1(c) in the main text is $\rho_0=\id_k\otimes\ket{\downarrow}\bra{\downarrow}$, uniformly distributed in quasimomentum space. When projecting $\rho_0(x)$ onto the quasimomentum basis,
we have $p_{kk'} = \bra{k}\rho_0(x)\ket{k'}$, which clearly depends on the width $d$.
Nevertheless, as shown in Fig.~\ref{figS5}(a) and (b), we observe that despite initializing different wave packets, the resulting deformed LTSTs $\widetilde{\langle \sigma_{x,y} (k_G)\rangle}$ in the GBZ, derived using Eq.~\eqref{LTST_measure}, remain nearly identical, which are also consistent with the results presented in Fig.~3(b) in the main text.
From the discussion above, it can be concluded that once the values of $p_{kk'}$ are determined by assuming a localized initial wave packet ($d\leq L/30$ in our calculations),
the dynamical field ${\bm g}(k_G)$ in the GBZ can always be constructed from observables measured in the BZ, enabling an accurate characterization of the non-Bloch topology.

Generally, the complex-momentum bases are not biorthogonal; however, we can still project the state onto these bases. 
To derive $p_{k_G k'_G}^{\sigma\sigma'}(t)$, we utilize the result
\begin{equation}~\label{rhot_Pkkss}
	\langle\beta^{\rm L}(k_G),\sigma| \rho(t) |\beta^{\rm L}(k'_G),\sigma'\rangle = \sum_{k''_G,k'''_G} p_{k''_Gk'''_G}^{\sigma\sigma'} \langle\beta^{\rm L}(k_G)
	 |\beta^{\rm R}(k''_G)\rangle \langle \beta^{\rm R}(k'''_G)|\beta^{\rm L}(k'_G)\rangle,
\end{equation}
where the left-hand side can be derived as
\begin{equation}
\langle\beta^{\rm L}(k_G),\sigma| \rho(t) |\beta^{\rm L}(k'_G),\sigma'\rangle=\sum_{k,k'}T_{k_Gk}p^{\sigma\sigma'}_{kk'}(t)T^*_{k_Gk'},
\end{equation}
with $T_{k_Gk}$ and $p^{\sigma\sigma'}_{kk'}$ defined in Eq.~\eqref{Tkk_def} and Eq.~\eqref{Psskk}, respectively.
For each pair $\{\sigma,\sigma'\}$,  Eq.~\eqref{rhot_Pkkss} corresponds to $N^2$ equations  ($N$ being the number of lattice sites), whose $N^2$ solutions yield $p_{k_Gk'_G}^{\sigma\sigma'}$.




\begin{thebibliography}{99}
\bibitem{footnote1} There is no constraint on the choice of $\rho_0$ for 1D systems with complex spectra~\cite{Suppl}. 
For instance, in the 1D NH SSH model, the initial state can be arbitrary. In our calculations, it is specifically chosen to be fully polarized along the $\sigma_z$ axis.
Furthermore, unless otherwise stated, the initial state is assumed to be uniformly distributed in quasimomentum space and explicitly represented as a spin state for simplicity.
\bibitem{footnote2} The biorthogonality condition holds only when the GBZ is circular, as is the case in the 1D NH SSH model. However, the projection procedure can still be applied in the case of a non-circular GBZ, although it will be slightly more complex. For more details, please refer to \cite{Suppl}.
\bibitem{footnote3} The initial spatial distribution of the wave packet is not crucial for characterizing the topology, provided it remains well-separated from the lattice boundary.
\end{thebibliography}
\end{document}